\documentclass[fleqn,10pt]{wlscirep}
\usepackage[utf8]{inputenc}
\usepackage[T1]{fontenc}
\title{Genuine multipartite entanglement measures based on multi-party teleportation capability}

\author[1, 2]{Minjin Choi}
\author[3, 4, *]{Eunok Bae}
\author[3, 5, $\dagger$]{Soojoon Lee}
\affil[1]{Division of National Supercomputing, Korea Institute of Science and Technology Information, Daejeon 34141, Korea}
\affil[2]{Qunova Computing, Inc., Daejeon 34051, Korea}
\affil[3]{School of Computational Sciences, Korea Institute for Advanced Study, Seoul 02455, Korea}
\affil[4]{Department of Mathematics, Research Institute for Natural Sciences, Hanyang University, Seoul 04763, Korea}
\affil[5]{Department of Mathematics and Research Institute for Basic Sciences, Kyung Hee University, Seoul 02447, Korea}
\affil[*]{eobae84@gmail.com}
\affil[$\dagger$]{level@khu.ac.kr}

\begin{abstract}
Quantifying entanglement is vital to understand entanglement as a resource in quantum information processing, and many entanglement measures have been suggested for this purpose. 
When mathematically defining an entanglement measure, we should consider the distinguishability between entangled and separable states, the invariance under local transformation, the monotonicity under local operations and classical communication, and the convexity. 
These are reasonable requirements but may be insufficient, in particular when taking into account the usefulness of quantum states in multi-party quantum information processing. 
Therefore, if we want to investigate multipartite entanglement as a resource, then it can be necessary to consider the usefulness of quantum states in multi-party quantum information processing when we define a multipartite entanglement measure.
In this paper, we define new multipartite entanglement measures for three-qubit systems based on the three-party teleportation capability, and show that these entanglement measures satisfy the requirements for being genuine multipartite entanglement measures. 
We also generalize our entanglement measures for $N$-qubit systems, where $N \ge 4$, and discuss that these quantities may be good candidates to measure genuine multipartite entanglement.

\end{abstract}
\begin{document}

\flushbottom
\maketitle
%
%
\thispagestyle{empty}


\section*{Introduction}
\label{sec:Introduction}
Entanglement is a crucial resource in quantum computing and quantum information tasks that cannot be explained classically.
Typical two-party quantum applications of entanglement as a resource include quantum teleportation~\cite{BBC93} and quantum key distribution~\cite{E91}, which are performed on bipartite entangled states.
However, there also exist multi-party quantum applications, such as multi-party quantum teleportation~\cite{KB98,HCN23}, conference key agreement~\cite{CL05}, and quantum secret sharing~\cite{HBB99}, where multipartite entanglement is considered as a resource.
In particular, genuine multipartite entanglement~(GME) is a significant concept for multipartite systems,
since it plays an essential role in quantum communication~\cite{KB98, YC06} and quantum cryptography~\cite{CL05, HBB99, DBW21}.
GME is also a critical resource in measurement-based quantum computing~\cite{BB09}, quantum-enhanced measurements~\cite{GLM04}, quantum phase transitions~\cite{ORO06, MA10}, and quantum spin chains~\cite{BDE05}.
Therefore, in order to make use of GME as a resource, its quantification is necessary. 

Entanglement measures are mathematical tools to quantify entanglement. 
For bipartite systems, concurrence is one of the well-known entanglement measures~\cite{BDS96, HW97, W98}.
It can distinguish between entangled and separable states and does not increase under local operations and classical communication~(LOCC), which are important requirements for quantifying entanglement.
For multipartite systems, it is more complicated to investigate entanglement, in particular GME. 
Even for three-qubit pure states, we should consider three different bipartition scenarios, $A|BC$, $B|AC$, and $C|AB$. 
In addition, there are two inequivalent classes, the Greenberger-Horne-Zeilinger~(GHZ) class and the W class, differentiated by stochastic LOCC~\cite{DVC00}.
A straightforward approach to define entanglement measures for quantifying GME is to deal with bipartite entanglement measures for all bipartitions.
For instance, the minimum and the geometric mean of concurrences for all bipartitions~\cite{MCC11,LS22} satisfy the conditions for being a GME measure~\cite{GT09,MCC11};
the distinguishability between genuinely multipartite entangled states and biseparable states, the invariance under local transformations, the monotonicity under LOCC, and the convexity.
The concurrence fill~\cite{XE21}, which is the square root of the area of the three-qubit concurrence triangle, was also proposed as a GME measure, but it has recently been shown that this measure does not satisfy the monotonicity under LOCC~\cite{GLC23}.

We now ask whether a GME measure can compare the usefulness of any pure states in some specific multipartite quantum information processing.
This question is natural when we use GME as a resource.
For example, a monotonic relationship exists between bipartite entanglement and teleportation fidelity for pure states~\cite{P94,HHH96}.
Indeed, such a concept of the \textit{proper} GME measure has been discussed~\cite{XE21},
 which makes the GHZ state rank as more entangled than the W state.  
This concept stems from the fact that the GHZ state can be more useful than the W state in three-qubit teleportation~\cite{JPO03}.
However, teleportation capabilities for other arbitrary pure states have not been taken into account.
In fact, the minimum and the geometric mean of concurrences for all bipartitions are \textit{proper} GME measures, but it is not difficult to find quantum states for which these GME measures and the three-qubit teleportation capability~\cite{LJK05} give different orders.

In order to appropriately utilize GME as a resource, we need a GME measure that can compare the usefulness of quantum states in a given quantum information processing.
In this paper, we first take account of three-qubit teleportation, and propose novel GME measures for three-qubit systems based on three-qubit teleportation capability.
To this end, we consider the maximal average teleportation fidelity of resulting states on the other parties obtained after a measurement by one of the parties, and prove that our measures based on the fidelity can be used to observe separability on three-qubit systems and does not increase on average under LOCC.
By comparing our GME measures with other GME measures, we show that there are quantum states such that their usefulness in three-qubit teleportation cannot be explained by the other GME measures, while it can naturally be done from ours.
We also show that our GME measures can be defined by using only two of the possible three fidelities, unlike the minimum and the geometric mean of concurrences should consider concurrences for all bipartitions.
In other words, we can make GME measures that have a simpler form. 

This paper is organized as follows.
We first introduce the maximal average teleportation fidelity obtained after one of the parties measures his/her system, and look into its properties.
After defining entanglement measures based on the three-qubit teleportation capability, we prove that they fulfill the conditions for the GME measures.
We give examples to show that our newly defined GME measures are more appropriate than the other GME measures to compare the capability of three-qubit teleportation.
We finally generalize our entanglement measures to $N$-qubit systems, and discuss that these $N$-partite entanglement measures have the potential to be GME measures by showing that GME is related to $N$-qubit teleportation capability when $N \ge 4$.


\section*{Main results}
\subsection*{Three-qubit teleportation capability and its properties}
\label{sec:TeleportationFidelity}
Three-qubit teleportation we consider proceeds as follows.
Suppose that three parties, Alice, Bob, and Charlie, share a three-qubit state.
After one performs an orthogonal measurement on his/her system,
the rest carry out the standard teleportation~\cite{BBC93} over the resulting state with the measurement outcome.
For instance, if the initial state is $\ket{\rm{GHZ}}_{ABC}=\frac{1}{\sqrt{2}}\left(\ket{000}_{ABC}+\ket{111}_{ABC}\right)$,
then having one of them measures his/her system in the $X$ basis $\left\{\ket{0_{x}}, \ket{1_{x}}\right\}$, 
where $\ket{t_{x}}=\frac{1}{\sqrt{2}}\left(\ket{0}+(-1)^{t}\ket{1}\right)$ for $t=0$ or $1$, 
makes it possible for them to perfectly accomplish three-qubit teleportation
since it can be written as
$\ket{\rm{GHZ}}_{ABC}=
\frac{1}{2}\left(\ket{0_{x}0_{x}0_{x}}_{ABC}+\ket{0_{x}1_{x}1_{x}}_{ABC}
+\ket{1_{x}0_{x}1_{x}}_{ABC}+\ket{1_{x}1_{x}0_{x}}_{ABC}\right)$.

Let us first look at the maximal fidelity of two-qubit teleportation.
For a given teleportation scheme $\Lambda_{\rho_{AB}}$ over a two-qubit state $\rho_{AB}$,
the fidelity of two-qubit teleportation is defined as~\cite{P94}
\begin{equation}
F\left(\Lambda_{\rho_{AB}}\right)=\int d\xi\bra{\xi}\Lambda_{\rho_{AB}}\left(\ket{\xi}\bra{\xi}\right)\ket{\xi}.
\end{equation}
It has been proven that 
when $\Lambda_{\rho_{AB}}$ represents the standard teleportation scheme over $\rho_{AB}$ to attain the maximal fidelity, 
the following equation holds~\cite{HHH99,BHH00}: 
\begin{equation}
F\left(\Lambda_{\rho_{AB}}\right)=\frac{2f\left(\rho_{AB}\right)+1}{3},
\end{equation}
where $f$ is the fully entangled fraction~\cite{BDS96},
which is given by
\begin{equation}
f\left(\rho_{AB}\right)=\rm{max}\bra{e}\rho_{AB}\ket{e},
\end{equation}
where the maximum is over all maximally entangled states $\ket{e}$ of two qubits.
The given state $\rho_{AB}$ is said to be useful for teleportation
if and only if $F(\Lambda_{\rho_{AB}})>2/3$~\cite{P94,MP95,HHH96}.

We now consider three-qubit teleportation capability.
Let $F_{ij}$ be the maximal average fidelity of teleportation over the resulting state in the systems $i$ and $j$ after measuring the system $k$, where $i$, $j$, and $k$ are distinct systems in $\{A, B, C\}$.
Since three-qubit teleportation consists of a one-qubit measurement and two-qubit teleportation, it is straightforwardly obtained that~\cite{LJK05}
\begin{equation}
\label{eq:tele_fidel}
F_{ij}\left(\rho_{ABC}\right)=\frac{2f_{ij}\left(\rho_{ABC}\right)+1}{3},
\end{equation}
where
\begin{equation}
\label{eq:tele_frac}
f_{ij}\left(\rho_{ABC}\right)=\max_{U_k}\sum_{t=0}^{1}
\bra{t}U_{k}\rho_{k}{U_{k}}^{\dagger}\ket{t}f\left(\rho_{ij}^{U_{k}, t}\right).
\end{equation}
Here, $U_{k}$ is a unitary operator,
that is, \{${U_{k}}^{\dagger}\ket{0},{U_{k}}^{\dagger}\ket{1}\}$ describes a one-qubit orthogonal measurement on the system $k$,
and $\rho_{ij}^{U_{k}, t}$ is the resulting state with the outcome $t$.
We say that a given state $\rho_{ABC}$ is useful for three-qubit teleportation if and only if $\min\{F_{AB}\left(\rho_{ABC}\right), F_{BC}\left(\rho_{ABC}\right), F_{CA}\left(\rho_{ABC}\right)\}>2/3$.

Before showing the properties of the maximal average teleportation fidelity $F_{ij}$, let us first consider the two-qubit maximal teleportation fidelity $F$.
Hereafter, we denote $\ket{\psi}_{S_{1}S_{2}} \in SEP(S_{1}:S_{2})$ when $\ket{\psi}_{S_{1}S_{2}}$ is separable between the systems $S_{1}$ and $S_{2}$ for simplicity.
By the Schmidt decomposition, any two-qubit pure state 
$\ket{\phi}_{AB}$ can be written in the form $\ket{\phi}_{AB}=\sqrt{a}\ket{u_{0}v_{0}}_{AB}+\sqrt{1-a}\ket{u_{1}v_{1}}_{AB}$, where $0 \le a \le 1$ and $\{\ket{u_{0}}, \ket{u_{1}}\}$, $\{\ket{v_{0}}, \ket{v_{1}}\}$ are orthonormal sets.
Thus, by calculating the entanglement fraction $f(\ket{\phi}_{AB})$,  we get $F\left(\Lambda_{\ket{\phi}_{AB}}\right)=\frac{2}{3}+\frac{2}{3}\sqrt{a(1-a)}$.
Note that the concurrence~\cite{HW97,W98} for a pure state $\ket{\psi}_{S_{1}S_{2}}$ is defined as
\begin{equation}
C\left(\ket{\psi}_{S_{1}S_{2}}\right)=\sqrt{2\left(1-\mathrm{Tr}\left(\varrho_{S_{1}}^{2}\right)\right)},
\end{equation}
where $\varrho_{S_{1}}$ is the reduced density operator of $\ket{\psi}_{S_{1}S_{2}}$,
so we have 
\begin{equation}
C(\ket{\phi}_{AB})=3F\left(\Lambda_{\ket{\phi}_{AB}}\right)-2.
\end{equation}
From this equation, we can see that $\ket{\phi}_{AB} \in SEP(A:B)$ if and only if $F\left(\Lambda_{\ket{\phi}_{AB}}\right)=\frac{2}{3}$,
and for two-qubit pure states, the more entangled state with respect to the concurrence, the higher the maximal teleportation fidelity $F$.
Moreover, since the concurrence satisfies the monotonicity under LOCC on pure states, so does the maximal teleportation fidelity $F$.

We now show that the maximal average teleportation fidelity $F_{ij}$ on three-qubit pure states has similar properties.
For three-qubit pure states, 
it has been shown that the following equation holds~\cite{LJK05}:
\begin{equation}
\label{eq:fidel_three}
F_{ij}\left(\ket{\phi}_{ABC}\right)=\frac{\sqrt{\left(\tau+C_{ij}^{2}\right)\left(\ket{\phi}_{ABC}\right)}+2}{3},
\end{equation}
where $\tau$ is the three-tangle~\cite{CKW00} 
and $C_{ij}$ is the concurrence for the reduced density operator $\rho_{ij}$ of $\ket{\phi}_{ABC}$.
We note that the three-tangle $\tau$ satisfies
\begin{equation}
\label{tangle}
\tau = C^{2}_{i(jk)}-C^{2}_{ij}-C^{2}_{ik}
\end{equation}
for any distinct $i$, $j$, and $k$, where $C_{i(jk)}$ denotes the concurrence between $i$ and the other system $jk$. 
For mixed states, the concurrence is defined by means of the convex roof extension. 
In particular, for two-qubit systems, the concurrence of mixed state can be computed by~\cite{W98}
\begin{equation}
C(\rho)=\mathrm{max}\{0, \lambda_{1}-\lambda_{2}-\lambda_{3}-\lambda_{4}\},
\end{equation}
where the $\lambda_{l}$s are eigenvalues of the matrix $X=\sqrt{\sqrt{\rho}(\sigma_{y}\otimes\sigma_{y})\rho^{*}(\sigma_{y}\otimes\sigma_{y})\sqrt{\rho}}$ in decreasing order, $\sigma_{y}$ is the Pauli $Y$ operator, and $\rho^{*}$ is the conjugate of $\rho$.
Hence, it is not difficult to calculate $F_{ij}$ for a given three-qubit pure state.

By using Eq.~(\ref{eq:fidel_three}), we obtain the following lemmas,
which are important properties when we define our GME measures. 
\begin{Lem}
\label{Lem1}
Let $\ket{\phi}_{ABC}$ be a three-qubit pure state.
Then for any distinct $i$, $j$, and $k$ in $\{A, B, C\}$,
$F_{ij}\left(\ket{\phi}_{ABC}\right)=\frac{2}{3}$ if and only if $\ket{\phi}_{ABC} \in SEP(i:jk)$ or $\ket{\phi}_{ABC} \in SEP(j:ik)$.
Moreover, if $F_{ij}\left(\ket{\phi}_{ABC}\right)>\frac{2}{3}$ and $F_{ik}\left(\ket{\phi}_{ABC}\right)>\frac{2}{3}$, then $F_{jk}\left(\ket{\phi}_{ABC}\right)>\frac{2}{3}$.
\end{Lem}

\begin{Lem}
\label{Lem2}
For three-qubit pure states, the maximal average teleportation fidelities $F_{AB}$, $F_{BC}$, and $F_{CA}$ does not increase on average under LOCC. 
\end{Lem}

We remark that Lemma~\ref{Lem1} and Lemma~\ref{Lem2} are not directly derived from the properties of the three-tangle $\tau$ and the concurrence $C_{ij}$ although we use Eq.~(\ref{eq:fidel_three}) to prove them.
If $\ket{\phi}_{ABC} \in SEP(i:jk)$ for some $i$, $j$, and $k$, then $\tau(\ket{\phi}_{ABC})=0$,
but the converse is not true.
For example, $\tau(\ket{W}_{ABC})=0$, where $\ket{W}_{ABC}=\frac{1}{\sqrt{3}}(\ket{001}_{ABC}+\ket{010}_{ABC}+\ket{100}_{ABC})$.
However, $\ket{W}_{ABC} \notin SEP(i:jk)$ for any distinct $i$, $j$ and $k$.
In addition, the concurrence $C_{ij}$ can increase under LOCC on three-qubit states.
Indeed, $C_{ij}(\ket{\rm{GHZ}})=0$, but after measuring the system $k$ in the $X$ basis, $C_{ij}$ of the resulting state becomes $1$.

\subsection*{GME measures based on three-qubit teleportation capability}
Let $\ket{\psi} \in \mathcal{H}_{1} \otimes \mathcal{H}_{2} \otimes \cdots \otimes \mathcal{H}_{N}$ be an $N$-partite pure state.
The state $\ket{\psi}$ is called biseparable if it can be written as $\ket{\psi}=\ket{\phi_{G}} \otimes \ket{\phi_{\bar{G}}}$, where $\ket{\phi_{G}} \in \mathcal{H}_{j_{1}} \otimes \cdots \otimes \mathcal{H}_{j_{k}}$ and $\ket{\phi_{\bar{G}}} \in \mathcal{H}_{j_{k+1}} \otimes \cdots \otimes \mathcal{H}_{j_{N}}$.
Here, $k<N$ and $\{j_{1},...,j_{k}|j_{k+1},...,j_{N}\}$ is a bipartition of the whole system.
An $N$-partite mixed state $\rho$ is called biseparable if it can be written as a convex sum of biseparable pure states $\rho=\sum_{i}p_{i}\ket{\psi_{i}}\bra{\psi_{i}}$, where the biseparable pure states $\{\ket{\psi_{i}}\}$ can be biseparable with respect to different bipartitions.
If an $N$-partite state is not biseparable, then it is a genuinely $N$-partite entangled state.

Note that minimal conditions for being a good entanglement measure have been suggested as follows~\cite{GT09,MCC11}:
\begin{itemize}
\item[(i)] $E(\rho)>0$ if and only if $\rho$ is a nonbiseparable state.
\item[(ii)] $E$ is invariant under local unitary transformations.
\item[(iii)] $E$ is not increasing on average under LOCC. That is, if we have states $\{\rho_{k}\}$ with probabilities $\{p_{k}\}$ after applying a LOCC transformation to $\rho$,
then $\sum_{k}p_{k}E(\rho_{k}) \le E(\rho)$.
\item[(iv)] $E$ is convex.
\end{itemize}
If a multipartite entanglement measure satisfies these conditions, then we call it a GME measure.
Our approach is to define a multipartite entanglement measure on pure states and generalize it to mixed states through the convex roof extension
\begin{equation}
E(\rho)=\min_{\{p_{l}, \psi_{l}\}}\sum_{l}p_{l}E\left(\ket{\psi_{l}}\right),
\end{equation}
where the minimum is over all possible decompositions $\rho=\sum_{l}p_{l}\ket{\psi_{l}}\bra{\psi_{l}}$.
This approach has the advantage that it suffices to define an entanglement measure on pure states that satisfies the conditions (i), (ii), and (iii) in order to construct a GME measure.

Let us now define entanglement measures based on the three-qubit teleportation fidelity.
\begin{Def}
\label{def3}
For a three-qubit pure state $\ket{\phi}_{ABC}$, let $\mathcal{T}_{ij}\left(\ket{\phi}_{ABC}\right)=3F_{ij}\left(\ket{\phi}_{ABC}\right)-2$, where $F_{ij}$ is the maximal average teleportation fidelity in Eq.~(\ref{eq:fidel_three}).
We define multipartite entanglement measures $\mathcal{T}_{min}$ and $\mathcal{T}_{GM}$ as
\begin{eqnarray}
&&\mathcal{T}_{min} \equiv \min\{\mathcal{T}_{AB}, \mathcal{T}_{BC}, \mathcal{T}_{CA}\},\nonumber \\
&&\mathcal{T}_{GM} \equiv \sqrt[3]{\mathcal{T}_{AB}\mathcal{T}_{BC}\mathcal{T}_{CA}},
\end{eqnarray}
respectively, on three-qubit pure states.
For three-qubit mixed states, we generalize them via the convex roof extension.
\end{Def}
The reason why we use $\mathcal{T}_{ij}$ instead of $F_{ij}$ itself is to set the values of $\mathcal{T}_{min}$ and $\mathcal{T}_{GM}$ between 0 and 1.
It directly follows from Lemma~\ref{Lem1} that $\mathcal{T}_{min}$ and $\mathcal{T}_{GM}$ satisfy the condition (i). 
We know that they are invariant under local transformations, which is the condition (ii), from the definition of $F_{ij}$.
From Lemma~\ref{Lem2}, we can also prove that they fulfill the condition (iii).
The condition (iv) is guaranteed by the convex roof extension.
Therefore, we have the following theorem.
\begin{Thm}
\label{Thm4}
Entanglement measures $\mathcal{T}_{min}$ and $\mathcal{T}_{GM}$ are GME measures.
\end{Thm}

In Lemma~\ref{Lem1}, we also showed that for a three-qubit pure state $\ket{\phi}_{ABC}$, if $F_{ij}\left(\ket{\phi}_{ABC}\right)>\frac{2}{3}$ and $F_{ik}\left(\ket{\phi}_{ABC}\right)>\frac{2}{3}$, then $F_{jk}\left(\ket{\phi}_{ABC}\right)>\frac{2}{3}$ for any distinct $i$, $j$, and $k$ in $\{A, B, C\}$.
Therefore, only two quantities $\mathcal{T}_{ij}$ and $\mathcal{T}_{ik}$ are enough to define a GME measure.
\begin{Def}
For any distinct $i$, $j$, and $k$ in $\{A, B, C\}$, we define multipartite entanglement measures $\mathcal{T}_{min}^{(i)}$ and $\mathcal{T}_{GM}^{(i)}$ as
\begin{eqnarray}
&&\mathcal{T}_{min}^{(i)} \equiv \min\{\mathcal{T}_{ij}, \mathcal{T}_{ik}\},\nonumber\\
&&\mathcal{T}_{GM}^{(i)} \equiv \sqrt{\mathcal{T}_{ij}\mathcal{T}_{ik}},
\end{eqnarray}
on three-qubit pure states.
For three-qubit mixed states, we generalize them through the convex roof extension.
\end{Def}
We have the following theorem by applying the same proof method in Theorem~\ref{Thm4}.
\begin{Thm}
\label{Thm6}
Entanglement measures $\mathcal{T}_{min}^{(i)}$ and $\mathcal{T}_{GM}^{(i)}$ are GME measures for any $i\in \{A,B,C\}$.
\end{Thm}
We can interpret entanglement measures $\mathcal{T}_{min}^{(i)}$ and $\mathcal{T}_{GM}^{(i)}$ as the minimum and the average teleportation capability of the system $i$, respectively.
Remark that if one defines an entanglement measure with concurrence in this way, then it cannot be a GME measure.
For example, let us think of the biseparable state
$\ket{\xi}_{ABC}=\frac{1}{\sqrt{2}}(\ket{000}_{ABC}+\ket{110}_{ABC}).$
We can clearly see that $C_{A(BC)}=C_{B(CA)}=1$, but $C_{C(AB)}=0$.
Thus, if we define $G_{min}\equiv\min\{C_{A(BC)}, C_{B(CA)}\}$, it cannot be a GME measure since $G_{min}(\ket{\xi}_{ABC}) \neq 0$.

The following examples show that our GME measures are more suitable to capture the usefulness of a given state for three-qubit teleportation.
We note that GME measures $C_{min}$ and $C_{GM}$ in References~\cite{MCC11,LS22} are given by $C_{min} \equiv \min\{C_{A(BC)}, C_{B(CA)}, C_{C(AB)}\}$ and $C_{GM} \equiv \sqrt[3]{C_{A(BC)}C_{B(CA)}C_{C(AB)}}$, respectively, on three-qubit pure states.

\begin{figure}[ht]
\label{fig_ex1}
\begin{center}
\includegraphics[width=13cm]{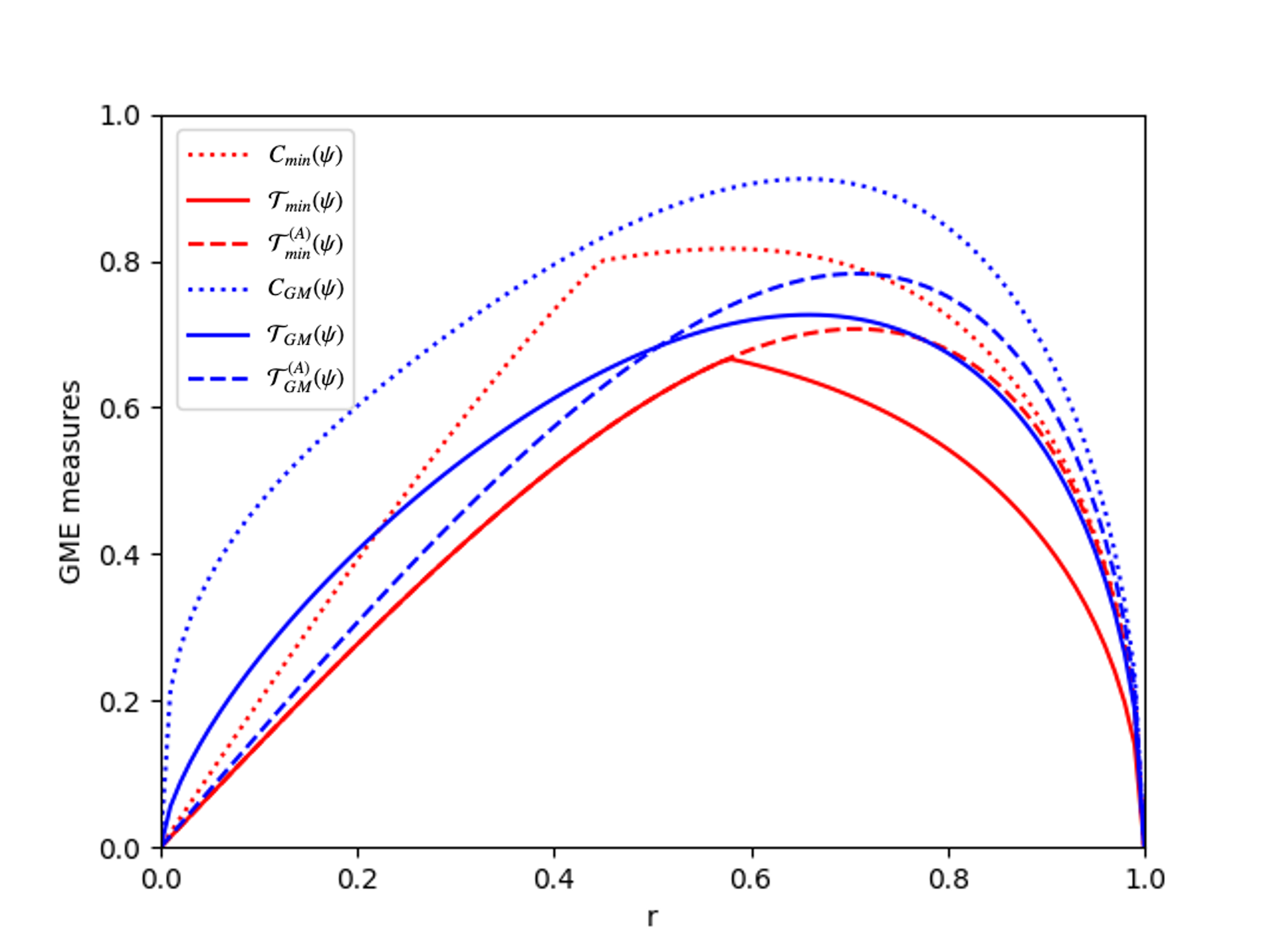}
\caption{Graphs of the GME measures $C_{min}$, $\mathcal{T}_{min}$, $\mathcal{T}^{(A)}_{min}$, $C_{GM}$, $\mathcal{T}_{GM}$, and $\mathcal{T}^{(A)}_{GM}$ for the state $\ket{\psi(r)}$ of the form in Eq.~(\ref{eq:ex_psi}).
For any two of these measures, we can choose $r$ for which these measures have different values.
Note that for the state $\ket{\phi(t)}$ of the form in Eq~(\ref{eq:ex_phi}), these measures have the same value $2t\sqrt{1-t^{2}}$ for any $t \in [0, 1]$, and these values vary from $0$ to $1$.
Hence, we can easily find states in which the values of the measures have different orders.
For instance, let $t'$ be a value such that $C_{GM}(\phi(t'))=\mathcal{T}_{GM}(\phi(t'))=0.8$.
Then $C_{GM}(\psi(0.7))>C_{GM}(\phi(t'))$, but $\mathcal{T}_{GM}(\psi(0.7))<\mathcal{T}_{GM}(\phi(t'))$. This means that $C_{GM}$ and $\mathcal{T}_{GM}$ rank the states $\psi(0.7)$ and $\phi(t')$ differently.}
\end{center}
\end{figure}

\begin{example}
\label{ex1}
We calculate $\mathcal{T}_{min}$, $\mathcal{T}_{GM}$, $\mathcal{T}^{(A)}_{min}$, $\mathcal{T}^{(A)}_{GM}$, $C_{min}$, and $C_{GM}$ for some states,
and show that they give the opposite order for the states, which means that these GME measures are distinct from one another.
For $0 \le t \le 1$, let
\begin{equation}
\label{eq:ex_phi}
\ket{\phi(t)}_{ABC}=t\ket{000}_{ABC}+\sqrt{1-t^{2}}\ket{111}_{ABC}.
\end{equation}
Then all GME measures return the same value $2t\sqrt{1-t^{2}}$ on the state $\ket{\phi(t)}_{ABC}$.
Note that $h(t) \equiv 2t\sqrt{1-t^{2}}$ is a continuous function of $t$, which has the minimum value $0$ at $t=0,1$ and the maximum value $1$ at $t=1/\sqrt{2}$.
Hence, if there is a state $\ket{\psi}_{ABC}$
such that $E\left(\psi\right)>E'\left(\psi\right)$,
where $E,E' \in \left\{\mathcal{T}_{min}, \mathcal{T}_{GM}, \mathcal{T}^{(A)}_{min}, \mathcal{T}^{(A)}_{GM}, C_{min}, C_{GM}\right\}$,
then it follows from the intermediate value theorem that there exists $t'$ with 
\begin{equation}
E\left(\psi\right)>E\left( \phi(t')\right)=
E'\left(\phi(t')\right)>E'\left(\psi\right),
\end{equation}
which means that these GME measures $E$ and $E'$ provide the opposite order for the quantum states $\ket{\psi}_{ABC}$ and $\ket{\phi(t')}_{ABC}$.
Indeed,
let us consider the following state
\begin{equation}
\label{eq:ex_psi}
\ket{\psi(r)}_{ABC} = r\ket{000}_{ABC} + \frac{\sqrt{1-r^2}}{2}\ket{101}_{ABC} + \frac{\sqrt{1-r^2}}{\sqrt{2}}\ket{110}_{ABC} + \frac{\sqrt{1-r^2}}{2}\ket{111}_{ABC},
\end{equation}
where $0 \le r \le 1$. 
Then for any $E,E' \in \left\{\mathcal{T}_{min}, \mathcal{T}_{GM}, \mathcal{T}^{(A)}_{min}, \mathcal{T}^{(A)}_{GM}, C_{min}, C_{GM}\right\}$, it is easy to find $r'$ such that $E\left(\psi(r')\right)>E'\left(\psi(r')\right)$
as seen in FIG.~1.
Hence, they are all different GME measures.
\end{example}

\begin{figure}[ht]
\begin{center}
\label{Fig2}
\includegraphics[width=13cm]{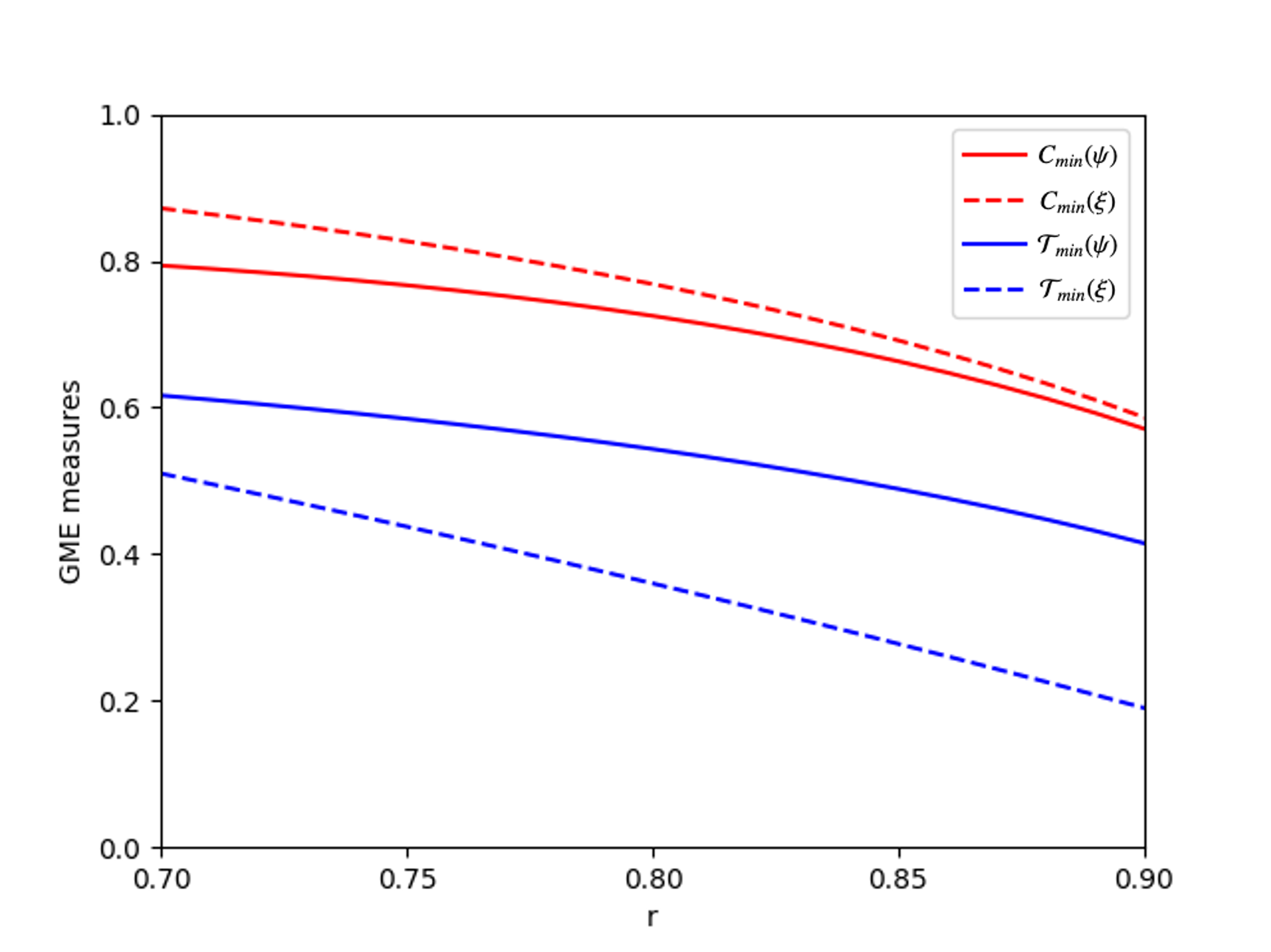}
\caption{If we calculate GME measures $C_{min}$ and $\mathcal{T}_{min}$ for the states $\ket{\psi(r)}$ in Eq.~(\ref{eq:ex_psi}) and $\ket{\xi(r)}$ in Eq.~(\ref{eq:ex_xi}), then $C_{min}\left(\xi(r)\right) > C_{\min}\left(\psi(r)\right)$ but $\mathcal{T}_{min}\left(\xi(r)\right)<\mathcal{T}_{min}\left(\psi(r)\right)$ for $0.7 \le r \le 0.9$.
Hence, we can say that $C_{min}$ is not appropriate for comparing teleportation capabilites.}
\end{center}
\end{figure}

\begin{example}
The GME measure $\mathcal{T}_{min}$ is defined based on three-qubit teleportation capability.
Thus, we can say that if $\ket{\psi}$ is more entangled than $\ket{\xi}$ with respect to $\mathcal{T}_{min}$, then $\ket{\psi}$ is more useful than $\ket{\xi}$ in three-qubit teleportation.
Let
\begin{equation}
\label{eq:ex_xi}
\ket{\xi(r)}_{ABC}=\frac{\sqrt{1-r^{2}}}{\sqrt{2}}\ket{001}_{ABC}+\frac{\sqrt{1-r^{2}}}{\sqrt{2}}\ket{010}_{ABC}+r\ket{100}_{ABC},
\end{equation}
where $0 \le r \le 1$.
In FIG.~2, 
we can see that for $0.7 < r <0.9$,
\begin{eqnarray}
&&C_{min}\left(\xi(r)\right) > C_{\min}\left(\psi(r)\right), \nonumber \\
 &&\mathcal{T}_{min}\left(\xi(r)\right)<\mathcal{T}_{min}\left(\psi(r)\right),
 \end{eqnarray}
where $\ket{\psi(r)}_{ABC}$ is the state in Eq.~(\ref{eq:ex_psi}).
In other words, although $\ket{\psi(r)}_{ABC}$ is more valuable for three-qubit teleportation than $\ket{\xi(r)}_{ABC}$ in this case, $C_{min}$ does not catch this fact.
Similar examples can be readily found for other GME measures as well.
\end{example}

\subsection*{GME and $N$-qubit teleportation capability}

We here discuss the relation between GME and $N$-qubit teleportation capability, where $N \ge 4$.
Note that $f_{ij}$ in Eq.~(\ref{eq:tele_frac}) can be generalized in the following two ways.
Let $\rho_{A_{1} \cdots A_{N}}$ be a $N$-qubit quantum state and $K=\{k_{1},k_{2}, \dots, k_{N-2}\}=\{A_{1},A_{2}, \dots, A_{N}\} \setminus \{i,j\}$.
One is
\begin{equation}
\label{eq:tele_frac_N_1}
f_{ij}^{(N)}(\rho_{A_{1} \cdots A_{N}})=\max_{U_{K}}\sum_{J \in \{0,1\}^{N-2}}
\bra{J}U_{K}\rho_{K}U_{K}^{\dagger}\ket{J}f\left(\rho^{U_{K},J}_{ij}\right),
\end{equation}
where $\rho_{K}=\rho_{k_{1}} \otimes \rho_{k_{2}} \otimes \cdots \otimes \rho_{k_{N-2}}$ and $U_{K}=U_{k_{1}} \otimes U_{k_{2}} \otimes \cdots \otimes U_{k_{N-2}}$ is a product of local unitary operators.
The other is
\begin{equation}
\label{eq:tele_frac_N_2}
\bar{f}_{ij}^{(N)}(\rho_{A_{1} \cdots A_{N}})=\max_{k_{l} \in K}\left(\max_{U_{k_{l}}}\sum_{t=0}^{1}
\bra{t}U_{k_{l}}\rho_{A_{k_{l}}}U_{k_{l}}^{\dagger}\ket{t}f_{ij}^{(N-1)}\left(\rho_{K_{l}}^{U_{k_{l}},t}\right)\right),
\end{equation}
where $K_{l}=K\setminus \{k_{l}\}$ and $f^{(3)}_{ij}=f_{ij}$ in Eq.~(\ref{eq:tele_frac}).
The difference between these two definitions is whether or not communication between assistants is allowed.
Hence, we obtain two different maximal average teleportation fidelities as follows:
\begin{eqnarray}
&& F_{ij}^{(N)}\left(\rho_{A_{1} \cdots A_{N}}\right) = \frac{2f_{ij}^{(N)}\left(\rho_{A_{1} \cdots A_{N}}\right)+1}{3}, \nonumber \\
&& \bar{F}_{ij}^{(N)}\left(\rho_{A_{1} \cdots A_{N}}\right) = \frac{2\bar{f}_{ij}^{(N)}\left(\rho_{A_{1} \cdots A_{N}}\right)+1}{3}.
\end{eqnarray}

Let us define new quantities based on these $N$-qubit teleportation capabilities with a similar way for three-qubit case as follows.
\begin{Def}
\label{def7}
For a $N$-qubit pure state $\ket{\phi}_{A_1\dots A_N}$, let $\mathcal{T}_{ij}^{(N)}\left(\ket{\phi}_{A_1\dots A_N}\right)=3F_{ij}^{(N)}\left(\ket{\phi}_{A_1\dots A_N}\right)-2$.
We define $\mathcal{T}_{min}^{(N)}$ and $\mathcal{T}_{GM}^{(N)}$ as
\begin{eqnarray}
\label{eq:def7}
&&\mathcal{T}_{min}^{(N)} \equiv \min_{i<j} \{\mathcal{T}_{ij}^{(N)}\},\nonumber \\
&&\mathcal{T}_{GM}^{(N)} \equiv \sqrt[m]{\prod_{i<j}\mathcal{T}_{ij}^{(N)}},
\end{eqnarray}
respectively, on $N$-qubit pure states, where $m=\binom{N}{2}$.
For $N$-qubit mixed states, we generalize them via the convex roof extension. In the same way, we define  $\bar{\mathcal{{T}}}_{min}^{(N)}$ and $\bar{\mathcal{T}}_{GM}^{(N)}$ by using $\bar{F}_{ij}^{(N)}$.
\end{Def}

Now, we discuss that the quantities proposed in Definition~\ref{def7} may be good candidates to be GME measures for $N$-qubit systems, where $N \ge 4$.
As we have already shown in Lemma~\ref{Lem1}, 
for a three-qubit pure state $\ket{\phi}_{ABC},$ 
$F_{ij}\left(\ket{\phi}_{ABC}\right)=\frac{2}{3}$ if and only if $\ket{\phi}_{ABC} \in SEP(i:jk)$ or $\ket{\phi}_{ABC} \in SEP(j:ik)$ for any distinct $i$, $j$, and $k$ in $\{A, B, C\}$. 
Here, we have a similar argument for an $N$-qubit pure state.

Let us now assume that for an $N$-qubit pure state $\ket{\phi}_{A_{1} \cdots A_{N}}$, there exists a bipartition $\{G_{i}|G_{j}\}$ of the whole system with $i \in G_{i}$ and $j \in G_{j}$ such that $\ket{\phi}_{A_{1} \cdots A_{N}} \in SEP(G_{i}:G_{j})$.
If $k \in G_{i}$, where $k \notin \{i, j\}$, then we clearly have that
the resulting state after performing an orthogonal measurement in the system $k$ is a pure state which belongs to $SEP(\tilde{G}_{i}:G_{j})$, where $\tilde{G}_{i}=G_{i}\setminus\{k\}$.
By continuing this process, we can see that the quantum state obtained after measuring all systems except $i$ and $j$ is a pure state and separable between systems $i$ and $j$.
Therefore, we have the following proposition.
\begin{Prop}
\label{prop7}
Let $\ket{\phi}_{A_{1} \cdots A_{N}}$ be an $N$-qubit pure state.
If $\ket{\phi}_{A_{1} \cdots A_{N}} \in SEP(G_{i}:G_{j})$ for a bipartition $\{G_{i}|G_{j}\}$ of the whole system with $i \in G_{i}$ and $j \in G_{j}$,
then $F_{ij}^{(N)}\left(\ket{\phi}_{A_{1} \cdots A_{N}}\right)=\bar{F}_{ij}^{(N)}\left(\ket{\phi}_{A_{1} \cdots A_{N}}\right)=\frac{2}{3}$.
\end{Prop}
We note that $F_{ij}^{(N)}\left(\ket{\phi}_{A_{1} \cdots A_{N}}\right) \le \bar{F}_{ij}^{(N)}\left(\ket{\phi}_{A_{1} \cdots A_{N}}\right)$ for any $N$. 
Therefore, $F_{ij}^{(N)}\left(\ket{\phi}_{A_{1} \cdots A_{N}}\right)>\frac{2}{3}$ for all distinct $i,j$ implies that $\ket{\phi}_{A_{1} \cdots A_{N}}$ is a genuinely $N$-partite entangled state.
However, we do not need to check all possible fidelities to verify that a given state is a genuinely $N$-partite entangled state.
Indeed, it suffices to check $\min_{j}\left\{F_{ij}^{(N)}\right\}>\frac{2}{3}$ for a fixed $i$ 
since if $\ket{\phi}_{A_{1} \cdots A_{N}} \in SEP(P:P')$ for a bipartition $\{P|P'\}$ of the whole system,
then we can always choose a system $k$ from the party that does not contain $i$ and $F_{ik}^{(N)}\left(\ket{\phi}_{A_{1} \cdots A_{N}}\right)=\frac{2}{3}$ by Proposition~\ref{prop7}.
For example, when $N=4$, if $F_{AB}^{(4)}\left(\ket{\phi}_{ABCD}\right)>\frac{2}{3}$, $F_{AC}^{(4)}\left(\ket{\phi}_{ABCD}\right)>\frac{2}{3}$ and $F_{AD}^{(4)}\left(\ket{\phi}_{ABCD}\right)>\frac{2}{3}$,
then the remaining fidelities are also greater than $\frac{2}{3}$ and so, $\ket{\phi}_{ABCD}$ is a genuinely quadripartite entangled state.

Conversely, does $F_{ij}^{(N)}\left(\ket{\phi}_{A_{1} \cdots A_{N}}\right)=\frac{2}{3}$ or $\bar{F}_{ij}^{(N)}\left(\ket{\phi}_{A_{1} \cdots A_{N}}\right)=\frac{2}{3}$ implies separability for a bipartition $\{G_{i}|G_{j}\}$ with $i \in G_{i}$ and $j \in G_{j}$?
If this holds for $F_{ij}^{(N)}$, then this also holds for $\bar{F}_{ij}^{(N)}$ since $F_{ij}^{(N)}\left(\ket{\phi}_{A_{1} \cdots A_{N}}\right) \le \bar{F}_{ij}^{(N)}\left(\ket{\phi}_{A_{1} \cdots A_{N}}\right)$ for any $N$.
For $N=4$, by checking all the cases, we get the following proposition.
\begin{Prop}
\label{prop8}
For $N=4$, if $F_{ij}^{(4)}\left(\ket{\phi}_{ABCD}\right)=\frac{2}{3}$, then $\ket{\phi}_{ABCD} \in SEP(G_{i}:G_{j})$ for some bipartition $\{G_{i}|G_{j}\}$ with $i \in G_{i}$ and $j \in G_{j}$.
\end{Prop}

To sum up, we can distinguish genuinely quadripartite entangled states and biseparable states by means of $F_{ij}^{(4)}$ or $\bar{F}_{ij}^{(4)}$.
By definition, $\mathcal{T}_{min}^{(4)}$, $\mathcal{T}_{GM}^{(4)}$, $\bar{\mathcal{{T}}}_{min}^{(4)}$, and $\bar{\mathcal{T}}_{GM}^{(4)}$ satisfy the GME condition (ii) and (iv), and hence, we have the following corollary.

\begin{Cor}
$\mathcal{T}_{min}^{(4)}$, $\mathcal{T}_{GM}^{(4)}$, $\bar{\mathcal{{T}}}_{min}^{(4)}$, and $\bar{\mathcal{T}}_{GM}^{(4)}$ satisfy the GME condition (i), (ii), and (iv).
\end{Cor}

Note that we expect to have the same argument with more complicated calculations when $N \ge 5$. Therefore, $\mathcal{T}_{min}^{(4)}$, $\mathcal{T}_{GM}^{(4)}$, $\bar{\mathcal{{T}}}_{min}^{(4)}$, and $\bar{\mathcal{T}}_{GM}^{(4)}$ have the potential to be GME measures.


\section*{Conclusion}
In this paper, we have introduced GME measures for three-qubit states based on three-qubit teleportation capability.
In order to do that, we have considered the maximal average teleportation fidelity of the resulting states on the other parties obtained after one of the parties measures his/her system.
We have shown that the fidelity can be used to observe separability, and does not increase on average under LOCC for three-qubit pure states,
and by using these properties, we have proven that our entanglement measures defined using the fidelity satisfy the conditions for the GME measure.

For three-qubit mixed states, we have defined our entanglement measures by means of the convex roof extension.
This method is a good way to make our entanglement measures satisfy the conditions for being a good entanglement measure on the mixed states, but it is not easy to get an exact value because it is defined as a minimum for all possible ensembles.
For a profound understanding of multipartite entanglement, it is necessary to find this value or at least find its lower bound.
Furthermore, it would be important to see how this value relates to multipartite teleportation capability.

We have shown that the maximal average fidelity of four-qubit teleportation can be used to distinguish genuinely quadripartite entangled states and biseparable states.
This could be generalized to $N$-qubit systems, where $N \ge 5$.
Hence, the quantities defined in Definition~\ref{def7} have the potential to be GME measures.
It is not easy to show that the entanglement measures satisfy the conditions for GME measures, in particular the monotonicity under LOCC, because no analytic form such as Eq.~(\ref{eq:fidel_three}) has been known for $N \ge 4$.
Our future work is to rigorously prove that these quantities are GME measures. 
Moreover, 
it would also be intriguing to explore entanglement measures for $N$-qudit systems using a similar approach since we can define $N$-qudit teleportation capability analogously as we define $N$-qubit teleportation capability in this work.

We note that there are other quantum information tasks that use GME, such as conference key agreement or secret sharing.
It has been shown that any multipartite private state, which is the general form of quantum state capable of conference key agreement, is a genuinely multipartite entangled state~\cite{DBW21}. 
Hence, it could be interesting to see if we can define GME measures based on those quantum information tasks.
Besides, there have been recently known quantum information tasks such as the quantum secure direct communication~\cite{ZS22} or controlled quantum teleportation based on quantum walks~\cite{SBZY23}. It would be also a possible future work to see how these tasks can be related to GME measures and how we define entanglement measures based on them.
\section*{Methods}

\subsection*{Proof of Lemma~\ref{Lem1}}
If $\ket{\phi}_{ABC} \in SEP(i:jk)$, which means $C_{i(jk)}(\ket{\phi}_{ABC})=0$,  
then we obtain $\left(\tau+C^{2}_{ij}\right)(\ket{\phi}_{ABC})=0$ and $\left(\tau+C^{2}_{ik}\right)(\ket{\phi}_{ABC})=0$ from Eq.~(\ref{tangle}).
In other words, $F_{ij}(\ket{\phi}_{ABC})=\frac{2}{3}$ and $F_{ik}(\ket{\phi}_{ABC})=\frac{2}{3}$.
Similarly, if $\ket{\phi}_{ABC} \in SEP(j:ik)$, we have $F_{ij}(\ket{\phi}_{ABC})=\frac{2}{3}$ and $F_{jk}(\ket{\phi}_{ABC})=\frac{2}{3}$. For both cases, $F_{ij}(\ket{\phi}_{ABC})=\frac{2}{3}$. 
Conversely, let us assume that $F_{ij}(\ket{\phi}_{ABC})=\frac{2}{3}$.
Then $\tau(\ket{\phi}_{ABC})=0$ and $C^{2}_{ij}(\ket{\phi}_{ABC})=0$ since both are nonnegative.
Note that any three-qubit pure state $\ket{\phi}_{ABC}$ can be written as~\cite{AAC00}
\begin{equation}
\label{shcmidt_three}
\ket{\phi}_{ABC}=\alpha_{0}\ket{000}_{ABC}+\alpha_{1}e^{\mathbf{i}\theta}\ket{100}_{ABC}+\alpha_{2}\ket{101}_{ABC}+\alpha_{3}\ket{110}_{ABC}+\alpha_{4}\ket{111}_{ABC},
\end{equation}
where $\mathbf{i}=\sqrt{-1}$, $0 \le \theta \le \pi$, $\alpha_{l} \ge 0$, and $\sum_{l}\alpha^{2}_{l}=1$.
Hence, it follows from straightforward calculations that
\begin{eqnarray}
\label{eq:phi}
&& \tau\left(\ket{\phi}_{ABC}\right) = 4\alpha_{0}^{2}\alpha_{4}^{2}, \nonumber \\
&& C^{2}_{AB}\left(\ket{\phi}_{ABC}\right) = 4\alpha_{0}^{2}\alpha_{3}^{2}, \nonumber \\
&& C^{2}_{BC}\left(\ket{\phi}_{ABC}\right) = 4\alpha_{1}^{2}\alpha_{4}^{2}+4\alpha_{2}^{2}\alpha_{3}^{2}-8\alpha_{1}\alpha_{2}\alpha_{3}\alpha_{4}\cos\theta, \nonumber \\
&& C^{2}_{CA}\left(\ket{\phi}_{ABC}\right) = 4\alpha_{0}^{2}\alpha_{2}^{2}.
\end{eqnarray}
From $\tau(\ket{\phi}_{ABC})=0$, we get $\alpha_{0}=0$ or $\alpha_{4}=0$.
If $\alpha_{0}=0$, then $C^{2}_{AB}(\ket{\phi}_{ABC})=C^{2}_{CA}(\ket{\phi}_{ABC})=0$,
and if $\alpha_{0} \neq 0$,
then $C^{2}_{BC}(\ket{\phi}_{ABC}) =0$ if and only if
$C^{2}_{AB}(\ket{\phi}_{ABC})=0$ or $C^{2}_{CA}(\ket{\phi}_{ABC})=0$.
Therefore, from Eq.~(\ref{tangle}), we can see that $F_{ij}(\ket{\phi}_{ABC})=\frac{2}{3}$ implies $C_{i(jk)}(\ket{\phi}_{ABC})=0$ or $C_{j(ki)}(\ket{\phi}_{ABC})=0$, that is,
$\ket{\phi}_{ABC} \in SEP(i:jk)$ or $\ket{\phi}_{ABC} \in SEP(j:ik)$.
Furthermore, we also have that if $F_{ij}(\ket{\phi}_{ABC})=\frac{2}{3}$, then $F_{ik}(\ket{\phi}_{ABC})=\frac{2}{3}$ or $F_{jk}(\ket{\phi}_{ABC})=\frac{2}{3}$.

\subsection*{Proof of Lemma~\ref{Lem2}}
It is sufficient to show that $\sqrt{\tau+C^{2}_{AB}}$, $\sqrt{\tau+C^{2}_{BC}}$, and $\sqrt{\tau+C^{2}_{CA}}$ do not increase on average under LOCC, thanks to Eq.~(\ref{eq:fidel_three}).
One important observation is that it is always possible to decompose any local protocol into positive operator-valued measures~(POVMs) such that only one party implements operations on his/her system. 
Moreover, we also remark that a generalized local POVM can be carried out by a sequence of POVMs with two outcomes~\cite{AJD00, DVC00}.
Without loss of generality, let us assume that Alice performs POVM consisting of two elements, say $A_{0}$ and $A_{1}$.
By using the singular value decomposition, they can be written as $A_{t}=U_{t}D_{t}V$, 
where $U_{t}$ and $V$ are unitary matrices, and $D_{t}$ are diagonal matrices
with entries $(a, b)$ and $\left(\sqrt{1-a^{2}}, \sqrt{1-b^{2}}\right)$, respectively, for some $a, b \in [0,1]$.
Here, the same unitary operation $V$ can be chosen for both $A_{0}$ and $A_{1}$ because they compose the POVM.

Let $\ket{\phi}_{ABC}$ be an initial state of the form in Eq.~(\ref{shcmidt_three}).
After Alice's POVM, we have $\ket{\psi_{t}}_{ABC}=A_{t}\ket{\phi}_{ABC}/\sqrt{p_{t}}$, where $p_{t}=\bra{\phi}A_{t}^{\dagger}A_{t}\ket{\phi}$.
Let us calculate the three-tangle and the concurrence of $\ket{\psi_{t}}_{ABC}$.
Since the three-tangle and the concurrence are invariant under local transformations,
it suffices to consider those of $D_{t}V\ket{\phi}_{ABC}/\sqrt{p_{t}}$ instead of $\ket{\psi_{t}}_{ABC}$ itself.
If we denote $v_{ij}=\bra{i}V\ket{j}$ and $d_{t,i}=\bra{i}D_{t}\ket{i}$ for $i,j,t \in \{0,1\}$, then
\begin{eqnarray}
D_{t}V\ket{\phi}_{ABC}=&d_{t,0}\left(v_{00}\alpha_{0}+v_{01}\alpha_{1}e^{\mathbf{i}\theta}\right)\ket{000}_{ABC}+d_{t,0}v_{01}\alpha_{2}\ket{001}_{ABC}+d_{t,0}v_{01}\alpha_{3}\ket{010}_{ABC}+d_{t,0}v_{01}\alpha_{4}\ket{011}_{ABC} \nonumber \\
&+d_{t,1}\left(v_{10}\alpha_{0}+v_{11}\alpha_{1}e^{\mathbf{i}\theta}\right)\ket{100}_{ABC}+d_{t,1}v_{11}\alpha_{2}\ket{101}_{ABC}+d_{t,1}v_{11}\alpha_{3}\ket{110}_{ABC}+d_{t,1}v_{11}\alpha_{4}\ket{111}_{ABC}.
\end{eqnarray}
By definition, with some tedious calculations, we can obtain
\begin{eqnarray}
&& \tau\left(\ket{\psi_{t}}_{ABC}\right)=\frac{4}{p_{t}^{2}}\cdot d_{t,0}^{2}d_{t,1}^{2}\alpha_{0}^{2}\alpha_{4}^{2}, \nonumber \\
&& C^{2}_{A(BC)}\left(\ket{\psi_{t}}_{ABC}\right)=\frac{4}{p_{t}^{2}}\cdot d_{t,0}^{2}d_{t,1}^{2}\alpha_{0}^{2}\left(\alpha_{2}^{2}+\alpha_{3}^{2}+\alpha_{4}^{2}\right), \nonumber \\
&& C^{2}_{B(CA)}\left(\ket{\psi_{t}}_{ABC}\right)=\frac{4}{p_{t}^{2}}\cdot \left(d_{t,0}^{2}d_{t,1}^{2}\alpha_{0}^{2}\alpha_{3}^{2}+g\left(\ket{\psi_{t}}_{ABC}\right)\right), \nonumber \\
&& C^{2}_{C(AB)}\left(\ket{\psi_{t}}_{ABC}\right)=\frac{4}{p_{t}^{2}}\cdot \left(d_{t,0}^{2}d_{t,1}^{2}\alpha_{0}^{2}\alpha_{2}^{2}+g\left(\ket{\psi_{t}}_{ABC}\right)\right),
\end{eqnarray}
where 
\begin{equation}
g\left(\ket{\psi_{t}}_{ABC}\right) \equiv \left(d_{t,0}^{2}|v_{01}|^{2}+d_{t,1}^{2}|v_{11}|^{2}\right) \cdot \sum_{i=0}^{1}d_{t,i}^{2}\left|\alpha_{4}\left(v_{i0}\alpha_{0}+v_{i1}\alpha_{1}e^{\mathbf{i}\theta}\right)-v_{i1}\alpha_{2}\alpha_{3}\right|^{2}
\end{equation}
From Eq.~(\ref{tangle}) and Eq.~(\ref{eq:phi}), we have
\begin{eqnarray}
&& \left(\tau+C^{2}_{AB}\right)\left(\ket{\psi_{t}}_{ABC}\right)=\frac{1}{p_{t}^{2}} \cdot d_{t,0}^{2}d_{t,1}^{2} \cdot \left(\tau+C^{2}_{AB}\right)\left(\ket{\phi}_{ABC}\right), \nonumber \\
&& \left(\tau+C^{2}_{BC}\right)\left(\ket{\psi_{t}}_{ABC}\right)=\frac{4}{p_{t}^{2}} \cdot g\left(\ket{\psi_{t}}_{ABC}\right), \nonumber \\
&& \left(\tau+C^{2}_{CA}\right)\left(\ket{\psi_{t}}_{ABC}\right)=\frac{1}{p_{t}^{2}} \cdot d_{t,0}^{2}d_{t,1}^{2} \cdot \left(\tau+C^{2}_{CA}\right)\left(\ket{\phi}_{ABC}\right).
\end{eqnarray}
One can readily check that 
\begin{equation}
\sqrt{\left(\tau+C^{2}_{ij}\right)\left(\ket{\phi}_{ABC}\right)} \ge \sum_{t=0}^{1}p_{t}\sqrt{\left(\tau+C^{2}_{ij}\right)\left(\ket{\psi_{t}}_{ABC}\right)},
\end{equation}
for $ij \in \{AB, CA\}$, since $\sum_{t=0}^{1} d_{t,0}d_{t,1} = ab+\sqrt{(1-a^{2})(1-b^{2})} \le 1$.
Hence, $\sqrt{\tau+C^{2}_{AB}}$ and $\sqrt{\tau+C^{2}_{CA}}$ do not increase on average under LOCC for three-qubit pure states.
Now, it remains to show that $\sqrt{\tau+C^{2}_{BC}}$ does not increase on average under LOCC for three-qubit pure states, or equivalently,
\begin{equation}
\label{eq:C_BC}
\sqrt{\left(\tau+C^{2}_{BC}\right)\left(\ket{\phi}_{ABC}\right)} \ge \sum_{t=0}^{1}p_{t}\sqrt{\left(\tau+C^{2}_{BC}\right)\left(\ket{\psi_{t}}_{ABC}\right)} = \sum_{t=0}^{1} 2\sqrt{g\left(\ket{\psi_{t}}_{ABC}\right)}.
\end{equation}
Observe that it can be written as
\begin{equation}
\left|\alpha_{4}\left(v_{i0}\alpha_{0}+v_{i1}\alpha_{1}e^{\mathbf{i}\theta}\right)-v_{i1}\alpha_{2}\alpha_{3}\right|=\left|\left<w | v_{i}\right>\right|,
\end{equation}
where $\ket{w}=\alpha_{0}\alpha_{4}\ket{0}+(\alpha_{1}\alpha_{4}e^{-\mathbf{i}\theta}-\alpha_{2}\alpha_{3})\ket{1}$ is an unnormalized vector and $\ket{v_{i}}=v_{i0}\ket{0}+v_{i1}\ket{1}$ for $i= 0, 1$.
In addition, since $\ket{v_{0}}$ and $\ket{v_{1}}$ are orthonormal,
we get 
\begin{equation}
\left|\left<w | v_{0}\right>\right|^{2}+\left|\left<w | v_{1}\right>\right|^{2}=||\ket{w}||^{2}=\frac{1}{4}\left(\tau+C^{2}_{BC}\right)\left(\ket{\phi}_{ABC}\right).
\end{equation}
Hence, we can reduce the desired inequality in Eq.~(\ref{eq:C_BC}) to 
\begin{equation}
||\ket{w}|| \ge \sqrt{g\left(\ket{\psi_{0}}_{ABC}\right)} + \sqrt{g\left(\ket{\psi_{1}}_{ABC}\right)}.
\end{equation}
With simple algebra, it can be readily shown that this inequality is equivalent to $\left(T||\ket{w}||^{2}-S\right)^{2} \ge 0$, where $T=a^{2}|v_{01}|^{2}+b^{2}|v_{11}|^{2}$ and $S=a^{2}\left|\left<w | v_{0}\right>\right|^{2}+b^{2}\left|\left<w | v_{1}\right>\right|^{2}$.
Therefore, Eq.~(\ref{eq:C_BC}) always holds, and so $\sqrt{\tau+C^{2}_{BC}}$ does not increase on average under LOCC for three-qubit pure states.

\subsection*{Proof of Theorem~\ref{Thm4}}
We here provide a rigorous proof that $\mathcal{T}_{min}$ and $\mathcal{T}_{GM}$ satisfy the condition (iii).
As in the proof of Lemma~\ref{Lem2},
we assume that one party performs a two-outcome POVM.
If we have $\ket{\psi_{t}}_{ABC}$ with probability $p_{t}$ after applying the POVM on $\ket{\phi}_{ABC}$,
then it follows from Lemma~\ref{Lem2} that
\begin{equation}
\mathcal{T}_{ij}\left(\ket{\phi}_{ABC}\right) \ge \sum_{t=0}^{1}p_{t}\mathcal{T}_{ij}\left(\ket{\psi_{t}}_{ABC}\right),
\end{equation}
for any distinct $i$ and $j$ in $\{A,B,C\}$.
Without loss of generality, let $\mathcal{T}_{AB}$ take the minimum, that is, $\mathcal{T}_{min}\left(\ket{\phi}_{ABC}\right)=\mathcal{T}_{AB}\left(\ket{\phi}_{ABC}\right)$.
Then
\begin{equation}
\mathcal{T}_{min}\left(\ket{\phi}_{ABC}\right) \ge \sum_{t=0}^{1}p_{t}\mathcal{T}_{AB}\left(\ket{\psi_{t}}_{ABC}\right) \ge \sum_{t=0}^{1}p_{t}\mathcal{T}_{min}\left(\ket{\psi_{t}}_{ABC}\right).
\end{equation}
Thus, $\mathcal{T}_{min}$ does not increase on average under LOCC for three-qubit pure states.
For $\mathcal{T}_{GM}$, we have
\begin{eqnarray}
\label{eq:40}
\mathcal{T}_{GM}\left(\ket{\phi}_{ABC}\right) &\ge& \sqrt[3]{\sum_{t=0}^{1}p_{t}\mathcal{T}_{AB}\left(\ket{\psi_{t}}_{ABC}\right)\sum_{s=0}^{1}p_{s}\mathcal{T}_{BC}\left(\ket{\psi_{s}}_{ABC}\right)\sum_{r=0}^{1}p_{r}\mathcal{T}_{CA}\left(\ket{\psi_{r}}_{ABC}\right)} \nonumber\\ &\ge&
\sum_{t=0}^{1}\sqrt[3]{p_{t}\mathcal{T}_{AB}\left(\ket{\psi_{t}}_{ABC}\right)}\sqrt[3]{p_{t}\mathcal{T}_{BC}\left(\ket{\psi_{t}}_{ABC}\right)}\sqrt[3]{p_{t}\mathcal{T}_{CA}\left(\ket{\psi_{t}}_{ABC}\right)} \nonumber\\
&=&\sum_{t=0}^{1}p_{t}\mathcal{T}_{GM}\left(\ket{\psi_{t}}_{ABC}\right).
\end{eqnarray}
The second inequality in Eq.~(\ref{eq:40}) comes from Mahler's inequality.
Hence, $\mathcal{T}_{GM}$ also does not increase on average under LOCC for three-qubit pure states.

Now, let $\rho$ be a three-qubit mixed state 
and $\mathcal{T}$ represent $\mathcal{T}_{min}$ or $\mathcal{T}_{GM}$. Assume that an ensemble 
${\{p_{l}, \psi_{l}\}}$ of $\rho$ attains the minimum of $\mathcal{T}(\rho)$,
that is,
\begin{equation}
\mathcal{T}(\rho)
= \sum_{l}p_{l}\mathcal{T}\left(\ket{\psi_{l}}\right).
\end{equation}
If we obtain 
\begin{equation}
\ket{\phi_{l,t}}=\frac{A_{t}\ket{\psi_{l}}}{\|A_{t}\ket{\psi_{l}}\|}
\end{equation}
after a POVM,
it follows from the monotonicity of $\mathcal{T}$ on pure states that
\begin{equation}
\mathcal{T}(\rho)
\ge \sum_{l,t}p_{l}q_{l,t}\mathcal{T}\left(\ket{\phi_{l,t}}\right),
\end{equation}
where $q_{l,t}=\|A_{t}\ket{\psi_{l}}\|^{2}$.
From the linearity of POVM, we note that 
the resulting state is
\begin{equation}
\sigma_{t}=\frac{1}{r_{t}}\sum_{l}p_{l}q_{l,t}\ket{\phi_{l,t}}\bra{\phi_{l,t}}
\end{equation}
with probability $r_{t}=\sum_{l}p_{l}q_{l,t}$.
By the definition of the convex roof extension, we finally have
\begin{equation}
\mathcal{T}(\sigma_{t}) \le \frac{1}{r_{t}}\sum_{l}p_{l}q_{l,t}\mathcal{T}\left(\ket{\phi_{l,t}}\right),
\end{equation}
which completes the proof.

\subsection*{Proof of Proposition~\ref{prop8}}
To prove the proposition, we first introduce the following lemma.
\begin{Lem}
\label{Lem9}
Let $\ket{\phi_{0}}$, $\ket{\phi_{1}}$, $\ket{\psi_{0}}$, and $\ket{\psi_{1}}$ be qubits and $|\inn{\phi_{0}}{\phi_{1}}| \neq 1$.
Define 
\begin{equation}
\ket{\xi}_{AB} \equiv 
\frac{1}{\sqrt{P}}\left(\alpha_{0}\ket{\phi_{0}}_{A}\ket{\psi_{0}}_{B} + \alpha_{1}\ket{\phi_{1}}_{A}\ket{\psi_{1}}_{B}\right),
\end{equation}
where $\alpha_{0}, \alpha_{1} \in \mathbb{C}$
with $|\alpha_{0}|^2+|\alpha_{1}|^2=1$, $|\alpha_{i}|\neq 0$ for $i=0, 1$, and $P \equiv 1 + 2Re\left(\alpha_{0}^{*}\alpha_{1}\inn{\phi_{0}}{\phi_{1}}\inn{\psi_{0}}{\psi_{1}}\right)$ is the normalization factor.
If $\ket{\xi}_{AB}$ is separable, then $\ket{\psi_{0}}$ is equivalent to $\ket{\psi_{1}}$ up to a global phase.
\end{Lem}
\subsubsection*{Proof of Lemma~\ref{Lem9}}
Since $\ket{\xi}_{AB}$ is separable, we have $\mathrm{Tr}\left(\sigma_{B}^{2}\right)=1$, where $\sigma_{B}$ is the reduced matrix of $\ket{\xi}_{AB}$.
Let us decompose $\ket{\phi_{1}}$ as $\ket{\phi_{1}}=\inn{\phi_{0}}{\phi_{1}}\ket{\phi_{0}}+\inn{\phi_{0}^{\perp}}{\phi_{1}}\ket{\phi_{0}^{\perp}}$,
where $\inn{\phi_{0}^{\perp}}{\phi_{0}}=0$.
Let $\beta = \inn{\phi_{0}}{\phi_{1}}$.
Then
\begin{equation}
P\sigma_{B} = |\alpha_{0}|^{2}\ket{\psi_{0}}\bra{\psi_{0}} + |\alpha_{1}|^{2}\ket{\psi_{1}}\bra{\psi_{1}}
+ \alpha_{0}^{*}\alpha_{1}\beta\ket{\psi_{1}}\bra{\psi_{0}}
+ \alpha_{0}\alpha_{1}^{*}\beta^{*}\ket{\psi_{0}}\bra{\psi_{1}}.
\end{equation}
By straightforward calculations, we have
\begin{eqnarray}
P^{2} &=& P^{2}\mathrm{Tr}\left(\sigma_{B}^{2}\right) \nonumber\\
&=& P^{2}-2|\alpha_{0}|^{2}|\alpha_{1}|^{2}(1-|\beta|^{2})(1-|\gamma|^{2}),
\end{eqnarray}
where $\gamma=\inn{\psi_{0}}{\psi_{1}}$.
Since $|\alpha_{i}| \neq 0$ for $i=0, 1$ and $|\beta| \neq 1$, we have $|\gamma|=1$.
Therefore, $\ket{\psi_{0}}$ is equivalent to $\ket{\psi_{1}}$ up to a global phase.

\subsubsection*{Proof of Proposition~\ref{prop8}}

Without loss of generality, let us assume that $F^{(4)}_{AB}(\ket{\phi}_{ABCD})=\frac{2}{3}$.
Then, it can be written as 
\begin{equation}
\ket{\phi}_{ABCD}=\alpha_{0}\ket{\eta_{0}}_{A}\ket{\zeta_{0}}_{B}\ket{00}_{CD}+\alpha_{1}\ket{\eta_{1}}_{A}\ket{\zeta_{1}}_{B}\ket{01}_{CD}+\alpha_{2}\ket{\eta_{2}}_{A}\ket{\zeta_{2}}_{B}\ket{10}_{CD}+\alpha_{3}\ket{\eta_{3}}_{A}\ket{\zeta_{3}}_{B}\ket{11}_{CD},
\end{equation}
where $\alpha_{i} \in \mathbb{C}$ with $\sum_{i=0}^{3}|\alpha_{i}|^{2}=1$.
If $|\alpha_{i}|=1$ for some $i$, then we clearly have $\ket{\phi}_{ABCD} \in SEP(G_{A}:G_{B})$ for any bipartition $\{G_{A}|G_{B}\}$.

If $|\alpha_{i}|^{2}+|\alpha_{j}|^{2}=1$ and $\alpha_{i}\alpha_{j} \ne 0$, where $(i, j) \in \{(0, 1), (0,2), (1,3), (2,3)\}$, then it can be immediately applied the results for the three-qubit system.
For example, let $|\alpha_{0}|^{2}+|\alpha_{2}|^{2}=1$,
then
$\ket{\phi}_{ABCD}=\ket{\phi'}_{ABC}\ket{0}_{D}$,
where $\ket{\phi'}_{ABC}=\alpha_{0}\ket{\eta_{0}}_{A}\ket{\zeta_{0}}_{B}\ket{0}_{C}+\alpha_{2}\ket{\eta_{2}}_{A}\ket{\zeta_{2}}_{B}\ket{1}_{C}$.
Since $F^{(4)}_{AB}(\ket{\phi}_{ABCD})=\frac{2}{3}$ and $\ket{\phi}_{ABCD} \in SEP(ABC:D)$,
we have $F^{(3)}_{AB}(\ket{\phi'}_{ABC})=\frac{2}{3}$.
Hence, $\ket{\phi'}_{ABC} \in SEP(A:BC)$ or $\ket{\phi'}_{ABC} \in SEP(AC:B)$,
and thus $\ket{\phi}_{ABCD} \in SEP(G_{A}:G_{B})$ for some bipartition $\{G_{A}|G_{B}\}$.
If $|\alpha_{0}|^{2}+|\alpha_{3}|^{2}=1$ and $\alpha_{0}\alpha_{3} \ne 0$, then 
\begin{equation}
\ket{\phi}_{ABCD}=\alpha_{0}\ket{\eta_{0}}_{A}\ket{\zeta_{0}}_{B}\ket{00}_{CD}+\alpha_{3}\ket{\eta_{3}}_{A}\ket{\zeta_{3}}_{B}\ket{11}_{CD}.
\end{equation}
In this case, after measuring each system $C$ and $D$ in the $X$ basis, we have  $\rho_{AB}^{H_{C}H_{D},J_{CD}=00}= 
\ket{\psi}_{AB}\bra{\psi}$
with $\ket{\psi}_{AB}=\frac{1}{\sqrt{P}}\left(\alpha_{0}\ket{\eta_{0}}_{A}\ket{\zeta_{0}}_{B} + \alpha_{3}\ket{\eta_{3}}_{A}\ket{\zeta_{3}}_{B}\right)$,
where $H$ is the Hadamard operator and $P$ is the normalization factor.
Since $\rho_{AB}^{H_{C}H_{D},J_{CD}=00}$ is separable due to the fact that $F^{(4)}_{AB}(\ket{\phi}_{ABCD})=\frac{2}{3}$, we apply Lemma~\ref{Lem9} to show separability of $\ket{\phi}$ between some bipartition $\{G_{A}|G_{B}\}$.
In case of $|\alpha_{1}|^{2}+|\alpha_{2}|^{2}=1$ with $\alpha_{1}\alpha_{2} \ne 0$, we can apply the same logic.

The next is the case of $|\alpha_{i}|^{2}+|\alpha_{j}|^{2}+|\alpha_{k}|^{2}=1$ with $\alpha_{i}\alpha_{j}\alpha_{k} \ne 0$.
For simplicity, $\ket{\phi} \sim \ket{\phi'}$ means $\ket{\phi}$ and $\ket{\phi'}$ are equivalent up to a global phase from now on.
By symmetry, it is enough to deal with the case of 
\begin{equation}
\ket{\phi}_{ABCD}=\alpha_{0}\ket{\eta_{0}}_{A}\ket{\zeta_{0}}_{B}\ket{00}_{CD}+\alpha_{1}\ket{\eta_{1}}_{A}\ket{\zeta_{1}}_{B}\ket{01}_{CD}+\alpha_{2}\ket{\eta_{2}}_{A}\ket{\zeta_{2}}_{B}\ket{10}_{CD}.
\end{equation}
Since $\rho_{AB}^{I_{C}H_{D},J_{CD}=00}$ and $\rho_{AB}^{H_{C}I_{D},J_{CD}=00}$ are separable, where $I$ is the identity operator, we have ($\ket{\eta_{0}} \sim \ket{\eta_{1}}$ or $\ket{\zeta_{0}} \sim \ket{\zeta_{1}}$) and ($\ket{\eta_{0}} \sim \ket{\eta_{2}}$ or $\ket{\zeta_{0}} \sim \ket{\zeta_{2}}$), respectively, by Lemma~\ref{Lem9}.
If $\ket{\eta_{0}} \sim \ket{\eta_{1}}$ and $\ket{\eta_{0}} \sim \ket{\eta_{2}}$,
then $\ket{\phi}_{ABCD} \in SEP(A:BCD)$.
Similarly, $\ket{\zeta_{0}} \sim \ket{\zeta_{1}}$ and $\ket{\zeta_{0}} \sim \ket{\zeta_{2}}$ 
implies $\ket{\phi}_{ABCD} \in SEP(ACD:B)$.
Let $\ket{\eta_{0}} \sim \ket{\eta_{1}}$ and $\ket{\zeta_{0}} \sim \ket{\zeta_{2}}$.
Then there exist real numbers $r_{1}$ and $s_{2}$ such that
$\ket{\eta_{0}}=e^{ir_{1}}\ket{\eta_{1}}$ and $\ket{\zeta_{0}}=e^{is_{2}}\ket{\zeta_{2}}$.
Hence, it can be written as
\begin{equation}
\ket{\phi}_{ABCD}=\alpha_{0}\ket{\eta_{0}}_{A}\ket{\zeta_{0}}_{B}\ket{00}_{CD}+\beta_{1}\ket{\eta_{0}}_{A}\ket{\zeta_{1}}_{B}\ket{01}_{CD}+\beta_{2}\ket{\eta_{2}}_{A}\ket{\zeta_{0}}_{B}\ket{10}_{CD},
\end{equation}
where $|\beta_{i}|=|\alpha_{i}|$ for $i=1,2$.
Note that 
$\rho_{AB}^{H_{C}H_{D},J_{CD}=00}=\ket{\psi}_{AB}\bra{\psi}$,
where
\begin{equation}
\ket{\psi}_{AB}=\frac{1}{\sqrt{P}}(\alpha_{0}\ket{\eta_{0}}_{A}\ket{\zeta_{0}}_{B}+\beta_{1}\ket{\eta_{0}}_{A}\ket{\zeta_{1}}_{B}+\beta_{2}\ket{\eta_{2}}_{A}\ket{\zeta_{0}}_{B})
\end{equation}
with the normalization factor $P$.
We can rewrite
\begin{equation}
\ket{\psi}_{AB}=\frac{1}{\sqrt{P'}}(\gamma_{0}\ket{\eta_{0}}_{A}\ket{\zeta'}_{B}+\gamma_{1}\ket{\eta_{2}}_{A}\ket{\zeta_{0}}_{B})
\end{equation}
with some nonzero values $\gamma_{0}, \gamma_{1} \in \mathbb{C}$ and $P' \in \mathbb{R}$.
By applying Lemma~\ref{Lem9} once again,
we have $\ket{\eta_{0}}\sim\ket{\eta_{2}}$ or $\ket{\zeta'}\sim\ket{\zeta_{0}}$.
In the former case, we get $\ket{\phi}_{ABCD} \in SEP(A:BCD)$.
In the latter case, $\ket{\zeta_{0}}\sim\ket{\zeta_{1}}$ since $\ket{\zeta'}$ is a linear combination of $\ket{\zeta_{0}}$ and $\ket{\zeta_{1}}$, and thus $\ket{\phi}_{ABCD} \in SEP(ACD:B)$.

Let us consider the case of $\alpha_{i} \ne 0$ for all $i$.
Since $\rho_{AB}^{I_{C}H_{D},J_{CD}=00}$, $\rho_{AB}^{H_{C}I_{D},J_{CD}=00}$, $\rho_{AB}^{H_{C}I_{D},J_{CD}=01}$, and $\rho_{AB}^{I_{C}H_{D},J_{CD}=10}$ are separable,
we have
\begin{eqnarray}
&&a_{1}: \ket{\eta_{0}} \sim \ket{\eta_{1}} ~~\rm{or}~~ b_{1}: \ket{\zeta_{0}} \sim \ket{\zeta_{1}}, \nonumber \\ 
&&a_{2}: \ket{\eta_{0}} \sim \ket{\eta_{2}} ~~\rm{or}~~ b_{2}: \ket{\zeta_{0}} \sim \ket{\zeta_{2}}, \nonumber \\
&&a_{3}: \ket{\eta_{1}} \sim \ket{\eta_{3}} ~~\rm{or}~~ b_{3}: \ket{\zeta_{1}} \sim \ket{\zeta_{3}}, \nonumber \\
&&a_{4}: \ket{\eta_{2}} \sim \ket{\eta_{3}} ~~\rm{or}~~ b_{4}: \ket{\zeta_{2}} \sim \ket{\zeta_{3}},
\end{eqnarray}
respectively.
Hence, there are 16 possible cases.

For a case in $\{a_{1}a_{2}a_{3}a_{4}, a_{1}a_{2}a_{3}b_{4}, a_{1}a_{2}b_{3}a_{4}, a_{1}b_{2}a_{3}a_{4}, b_{1}a_{2}a_{3}a_{4}\}$,
we have $\eta_{0} \sim \eta_{i}$ for all $i$, and so $\ket{\phi}_{ABCD} \in SEP(A:BCD)$.
If a case is in $\{b_{1}b_{2}b_{3}b_{4}, b_{1}b_{2}b_{3}a_{4}, b_{1}b_{2}a_{3}b_{4}, b_{1}a_{2}b_{3}b_{4}, a_{1}b_{2}b_{3}b_{4}\}$,
we obtain $\ket{\phi}_{ABCD} \in SEP(ACD:B)$.
For the case $a_{1}a_{2}b_{3}b_{4}$, we have $\ket{\eta_{0}} \sim \ket{\eta_{1}} \sim \ket{\eta_{2}}$ and $\ket{\zeta_{1}} \sim \ket{\zeta_{2}} \sim \ket{\zeta_{3}}$.
In this case, 
after applying the method used in the case of $|\alpha_{i}|^{2}+|\alpha_{j}|^{2}+|\alpha_{k}|^{2}=1$ with $\alpha_{i}\alpha_{j}\alpha_{k} \ne 0$ to $\rho_{AB}^{H_{C}H_{D},J_{CD}=00}$,
we can show separability.
For the cases $a_{1}b_{2}a_{3}b_{4}$, $b_{1}b_{2}a_{3}a_{4}$, and $b_{1}a_{2}b_{3}a_{4}$, it can be proven in the same way.

The remainder cases are $a_{1}b_{2}b_{3}a_{4}$ and $b_{1}a_{2}a_{3}b_{4}$. By symmetry, it suffices to consider the case $a_{1}b_{2}b_{3}a_{4}$.
In this case, the state can be written as
\begin{equation}
\ket{\phi}_{ABCD}=\beta_{0}\ket{\eta_{0}}_{A}\ket{\zeta_{0}}_{B}\ket{00}_{CD}+\beta_{1}\ket{\eta_{0}}_{A}\ket{\zeta_{1}}_{B}\ket{01}_{CD}+\beta_{2}\ket{\eta_{2}}_{A}\ket{\zeta_{0}}_{B}\ket{10}_{CD}+\beta_{3}\ket{\eta_{2}}_{A}\ket{\zeta_{1}}_{B}\ket{11}_{CD}
\end{equation}
for some $\beta_{i}\in\mathbb{C}$ with $|\alpha_{i}|=|\beta_{i}|$ for all $i$.
Then 
$\rho_{AB}^{H_{C}H_{D},J_{CD}=00}=\ket{\psi}_{AB}\bra{\psi},$
where
\begin{equation}
\ket{\psi}_{AB}=\frac{1}{\sqrt{P}}(\beta_{0}\ket{\eta_{0}}_{A}\ket{\zeta_{0}}_{B}+\beta_{1}\ket{\eta_{0}}_{A}\ket{\zeta_{1}}_{B}+\beta_{2}\ket{\eta_{2}}_{A}\ket{\zeta_{0}}_{B}+\beta_{3}\ket{\eta_{2}}_{A}\ket{\zeta_{1}}_{B})
\end{equation}
with a normalization factor $P$.
We can rewrite 
\begin{equation}
\ket{\psi}_{AB}=\frac{1}{\sqrt{P'}}(\gamma_{0}\ket{\eta_{0}}_{A}\ket{\zeta'}_{B}+\gamma_{2}\ket{\eta_{2}}_{A}\ket{\zeta''}_{B})
\end{equation}
with nonzero values $\gamma_{0}$, $\gamma_{2}$, and $P'$, where $\ket{\zeta'}$ and $\ket{\zeta''}$ are linear combinations of $\ket{\zeta_{0}}$ and $\ket{\zeta_{1}}$. 
Since $\rho_{AB}^{H_{C}H_{D},J_{CD}=00}$ is separable, by applying the Lemma~\ref{Lem9}, we have $\ket{\eta_{0}} \sim \ket{\eta_{2}}$ or $\ket{\zeta'} \sim \ket{\zeta''}$.
In the former case, we have $\ket{\phi}_{ABCD} \in SEP(A:BCD)$.
Let us consider the latter case.
We note that $\ket{\zeta'}=\delta_{0}\ket{\zeta_{0}}_{B}+\delta_{1}\ket{\zeta_{1}}_{B}$ and $\ket{\zeta''}=\delta_{2}\ket{\zeta_{0}}_{B}+\delta_{3}\ket{\zeta_{1}}_{B}$ for some $\delta_{i}$.
Since $\ket{\zeta'} \sim \ket{\zeta''}$, there exits $\theta$ such that
$\delta_{0}\ket{\zeta_{0}}_{B}+\delta_{1}\ket{\zeta_{1}}_{B} =e^{i\theta}( \delta_{2}\ket{\zeta_{0}}_{B}+\delta_{3}\ket{\zeta_{1}}_{B})$.
If $\zeta_{0} \sim \zeta_{1}$, then $\ket{\phi}_{ABCD} \in SEP(ACD:B)$.
In the other case, we have
$\delta_{0}=e^{i\theta}\delta_{2}$ and $\delta_{1}=e^{i\theta}\delta_{3}$ since $\ket{\zeta_{0}}$ and $\ket{\zeta_{1}}$ are linearly independent.
Hence, it can be rewritten as 
\begin{equation}
\ket{\phi}_{ABCD}=\tilde{\gamma}_{0}\delta_{0}\ket{\eta_{0}}_{A}\ket{\zeta_{0}}_{B}\ket{00}_{CD}+\tilde{\gamma}_{0}\delta_{1}\ket{\eta_{0}}_{A}\ket{\zeta_{1}}_{B}\ket{01}_{CD}+\tilde{\gamma}_{1}\delta_{0}\ket{\eta_{2}}_{A}\ket{\zeta_{0}}_{B}\ket{10}_{CD}+\tilde{\gamma}_{1}\delta_{1}\ket{\eta_{2}}_{A}\ket{\zeta_{1}}_{B}\ket{11}_{CD}
\end{equation}
with some complex values $\tilde{\gamma}_{i}$ and $\delta_{j}$.
Reordering the systems, we can have
\begin{equation}
(\tilde{\gamma}_{0}\ket{\eta_{0}}_{A}\ket{0}_{C}+\tilde{\gamma}_{1}\ket{\eta_{2}}_{A}\ket{1}_{C})(\delta_{0}\ket{\zeta_{0}}_{B}\ket{0}_{D}+\delta_{1}\ket{\zeta_{1}}_{B}\ket{1}_{D}),
\end{equation}
which means $\ket{\phi}_{ABCD} \in SEP(AC:BD)$.

\section*{Data availability}
No datasets were generated or analysed during the current study. The results were calculated by hand.
Correspondence and requests for materials should be addressed to E.B. or S.L..

\bibliography{reference}

\begin{thebibliography}{10}
\urlstyle{rm}
\expandafter\ifx\csname url\endcsname\relax
  \def\url#1{\texttt{#1}}\fi
\expandafter\ifx\csname urlprefix\endcsname\relax\def\urlprefix{URL }\fi
\expandafter\ifx\csname doiprefix\endcsname\relax\def\doiprefix{DOI: }\fi
\providecommand{\bibinfo}[2]{#2}
\providecommand{\eprint}[2][]{\url{#2}}

\bibitem{BBC93}
\bibinfo{author}{Bennett, C.~H.} \emph{et~al.}
\newblock \bibinfo{journal}{\bibinfo{title}{Teleporting an unknown quantum
  state via dual classical and einstein-podolsky-rosen channels}}.
\newblock {\emph{\JournalTitle{Phys. Rev. Lett.}}}
  \textbf{\bibinfo{volume}{70}}, \bibinfo{pages}{1895} (\bibinfo{year}{1993}).

\bibitem{E91}
\bibinfo{author}{Ekert, A.~K.}
\newblock \bibinfo{journal}{\bibinfo{title}{Quantum cryptography based on
  bell’s theorem}}.
\newblock {\emph{\JournalTitle{Phys. Rev. Lett.}}}
  \textbf{\bibinfo{volume}{67}}, \bibinfo{pages}{661} (\bibinfo{year}{1991}).

\bibitem{KB98}
\bibinfo{author}{Karlsson, A.} \& \bibinfo{author}{Bourennane, M.}
\newblock \bibinfo{journal}{\bibinfo{title}{Quantum teleportation using
  three-particle entanglement}}.
\newblock {\emph{\JournalTitle{Phys. Rev. A}}} \textbf{\bibinfo{volume}{58}},
  \bibinfo{pages}{4394} (\bibinfo{year}{1998}).

\bibitem{HCN23}
\bibinfo{author}{Harraz, S.}, \bibinfo{author}{Cong, S.} \&
  \bibinfo{author}{Nieto, J.~J.}
\newblock \bibinfo{journal}{\bibinfo{title}{Optimal tripartite quantum
  teleportation protocol through noisy channels}}.
\newblock {\emph{\JournalTitle{Quantum Information Processing}}}
  \textbf{\bibinfo{volume}{22}} (\bibinfo{year}{2023}).

\bibitem{CL05}
\bibinfo{author}{Chen, K.} \& \bibinfo{author}{Lo, H.-K.}
\newblock \bibinfo{title}{Conference key agreement and quantum sharing of
  classical secrets with noisy {GHZ} states}.
\newblock In \emph{\bibinfo{booktitle}{Proceedings. International Symposium on
  Information Theory, 2005. ISIT 2005.}}, \bibinfo{pages}{1607--1611}
  (\bibinfo{organization}{IEEE}, \bibinfo{year}{2005}).

\bibitem{HBB99}
\bibinfo{author}{Hillery, M.}, \bibinfo{author}{Bužek, V.} \&
  \bibinfo{author}{Berthiaume, A.}
\newblock \bibinfo{journal}{\bibinfo{title}{Quantum secret sharing}}.
\newblock {\emph{\JournalTitle{Phys. Rev. A}}} \textbf{\bibinfo{volume}{59}},
  \bibinfo{pages}{1829} (\bibinfo{year}{1999}).

\bibitem{YC06}
\bibinfo{author}{Yeo, Y.} \& \bibinfo{author}{Chua, W.~K.}
\newblock \bibinfo{journal}{\bibinfo{title}{Teleportation and dense coding with
  genuine multipartite entanglement}}.
\newblock {\emph{\JournalTitle{Phys. Rev. Lett.}}}
  \textbf{\bibinfo{volume}{96}}, \bibinfo{pages}{060502}
  (\bibinfo{year}{2006}).

\bibitem{DBW21}
\bibinfo{author}{Das, S.}, \bibinfo{author}{Bäuml, S.},
  \bibinfo{author}{Winczewski, M.} \& \bibinfo{author}{Horodecki, K.}
\newblock \bibinfo{journal}{\bibinfo{title}{Universal limitations on quantum
  key distribution over a network}}.
\newblock {\emph{\JournalTitle{Phys. Rev. X}}} \textbf{\bibinfo{volume}{11}},
  \bibinfo{pages}{041016} (\bibinfo{year}{2021}).

\bibitem{BB09}
\bibinfo{author}{Briegel, H.~J.}, \bibinfo{author}{Browne, D.~E.},
  \bibinfo{author}{D{\"u}r, W.}, \bibinfo{author}{Raussendorf, R.} \&
  \bibinfo{author}{Van~den Nest, M.}
\newblock \bibinfo{journal}{\bibinfo{title}{Measurement-based quantum
  computation}}.
\newblock {\emph{\JournalTitle{Nature Physics}}} \textbf{\bibinfo{volume}{5}},
  \bibinfo{pages}{19--26} (\bibinfo{year}{2009}).

\bibitem{GLM04}
\bibinfo{author}{Giovannetti, V.}, \bibinfo{author}{Lloyd, S.} \&
  \bibinfo{author}{Maccone, L.}
\newblock \bibinfo{journal}{\bibinfo{title}{Quantum-enhanced measurements:
  beating the standard quantum limit}}.
\newblock {\emph{\JournalTitle{Science}}} \textbf{\bibinfo{volume}{306}},
  \bibinfo{pages}{1330--1336} (\bibinfo{year}{2004}).

\bibitem{ORO06}
\bibinfo{author}{de~Oliveira, T.~R.}, \bibinfo{author}{Rigolin, G.} \&
  \bibinfo{author}{de~Oliveira, M.~C.}
\newblock \bibinfo{journal}{\bibinfo{title}{Genuine multipartite entanglement
  in quantum phase transitions}}.
\newblock {\emph{\JournalTitle{Physical Review A}}}
  \textbf{\bibinfo{volume}{73}}, \bibinfo{pages}{010305}
  (\bibinfo{year}{2006}).

\bibitem{MA10}
\bibinfo{author}{Montakhab, A.} \& \bibinfo{author}{Asadian, A.}
\newblock \bibinfo{journal}{\bibinfo{title}{Multipartite entanglement and
  quantum phase transitions in the one-, two-, and three-dimensional
  transverse-field ising model}}.
\newblock {\emph{\JournalTitle{Physical Review A}}}
  \textbf{\bibinfo{volume}{82}}, \bibinfo{pages}{062313}
  (\bibinfo{year}{2010}).

\bibitem{BDE05}
\bibinfo{author}{Bru{\ss}, D.}, \bibinfo{author}{Datta, N.},
  \bibinfo{author}{Ekert, A.}, \bibinfo{author}{Kwek, L.~C.} \&
  \bibinfo{author}{Macchiavello, C.}
\newblock \bibinfo{journal}{\bibinfo{title}{Multipartite entanglement in
  quantum spin chains}}.
\newblock {\emph{\JournalTitle{Physical Review A}}}
  \textbf{\bibinfo{volume}{72}}, \bibinfo{pages}{014301}
  (\bibinfo{year}{2005}).

\bibitem{BDS96}
\bibinfo{author}{Bennett, C.~H.}, \bibinfo{author}{DiVincenzo, D.~P.},
  \bibinfo{author}{Smolin, J.~A.} \& \bibinfo{author}{Wootters, W.~K.}
\newblock \bibinfo{journal}{\bibinfo{title}{Mixed-state entanglement and
  quantum error correction}}.
\newblock {\emph{\JournalTitle{Phys. Rev. A}}} \textbf{\bibinfo{volume}{54}},
  \bibinfo{pages}{3824} (\bibinfo{year}{1996}).

\bibitem{HW97}
\bibinfo{author}{Hill, S.~A.} \& \bibinfo{author}{Wootters, W.~K.}
\newblock \bibinfo{journal}{\bibinfo{title}{Entanglement of a pair of quantum
  bits}}.
\newblock {\emph{\JournalTitle{Phys. Rev. Lett.}}}
  \textbf{\bibinfo{volume}{78}}, \bibinfo{pages}{5022} (\bibinfo{year}{1997}).

\bibitem{W98}
\bibinfo{author}{Wootters, W.~K.}
\newblock \bibinfo{journal}{\bibinfo{title}{Entanglement of formation of an
  arbitrary state of two qubits}}.
\newblock {\emph{\JournalTitle{Phys. Rev. Lett.}}}
  \textbf{\bibinfo{volume}{80}}, \bibinfo{pages}{2245} (\bibinfo{year}{1998}).

\bibitem{DVC00}
\bibinfo{author}{Dür, W.}, \bibinfo{author}{Vidal, G.} \&
  \bibinfo{author}{Cirac, J.~I.}
\newblock \bibinfo{journal}{\bibinfo{title}{Three qubits can be entangled in
  two inequivalent ways}}.
\newblock {\emph{\JournalTitle{Phys. Rev. A}}} \textbf{\bibinfo{volume}{62}},
  \bibinfo{pages}{062314} (\bibinfo{year}{2000}).

\bibitem{MCC11}
\bibinfo{author}{Ma, Z.-H.} \emph{et~al.}
\newblock \bibinfo{journal}{\bibinfo{title}{Measure of genuine multipartite
  entanglement with computable lower bounds}}.
\newblock {\emph{\JournalTitle{Phys. Rev. A}}} \textbf{\bibinfo{volume}{83}},
  \bibinfo{pages}{062325} (\bibinfo{year}{2011}).

\bibitem{LS22}
\bibinfo{author}{Li, Y.} \& \bibinfo{author}{Shang, J.}
\newblock \bibinfo{journal}{\bibinfo{title}{Geometric mean of bipartite
  concurrences as a genuine multipartite entanglement measure}}.
\newblock {\emph{\JournalTitle{Phys. Rev. Research}}}
  \textbf{\bibinfo{volume}{4}}, \bibinfo{pages}{023059} (\bibinfo{year}{2022}).

\bibitem{GT09}
\bibinfo{author}{Gühne, O.} \& \bibinfo{author}{Tóth, G.}
\newblock \bibinfo{journal}{\bibinfo{title}{Entanglement detection}}.
\newblock {\emph{\JournalTitle{Phys. Rep.}}} \textbf{\bibinfo{volume}{474}},
  \bibinfo{pages}{1--75} (\bibinfo{year}{2009}).

\bibitem{XE21}
\bibinfo{author}{Xie, S.} \& \bibinfo{author}{Eberly, J.~H.}
\newblock \bibinfo{journal}{\bibinfo{title}{Triangle measure of tripartite
  entanglement}}.
\newblock {\emph{\JournalTitle{Phys. Rev. Lett.}}}
  \textbf{\bibinfo{volume}{127}}, \bibinfo{pages}{040403}
  (\bibinfo{year}{2021}).

\bibitem{GLC23}
\bibinfo{author}{Ge, X.}, \bibinfo{author}{Liu, L.} \& \bibinfo{author}{Cheng,
  S.}
\newblock \bibinfo{journal}{\bibinfo{title}{Tripartite entanglement measure
  under local operations and classical communication}}.
\newblock {\emph{\JournalTitle{Phys. Rev. A}}} \textbf{\bibinfo{volume}{107}},
  \bibinfo{pages}{032405} (\bibinfo{year}{2023}).

\bibitem{P94}
\bibinfo{author}{Popescu, S.}
\newblock \bibinfo{journal}{\bibinfo{title}{Bell’s inequalities versus
  teleportation: What is nonlocality?}}
\newblock {\emph{\JournalTitle{Phys. Rev. Lett.}}}
  \textbf{\bibinfo{volume}{72}}, \bibinfo{pages}{797} (\bibinfo{year}{1994}).

\bibitem{HHH96}
\bibinfo{author}{Horodecki, R.}, \bibinfo{author}{Horodecki, M.} \&
  \bibinfo{author}{Horodecki, P.}
\newblock \bibinfo{journal}{\bibinfo{title}{Teleportation, bell's inequalities
  and inseparability}}.
\newblock {\emph{\JournalTitle{Phys. Lett. A}}} \textbf{\bibinfo{volume}{222}},
  \bibinfo{pages}{21} (\bibinfo{year}{1996}).

\bibitem{JPO03}
\bibinfo{author}{Joo, J.}, \bibinfo{author}{Park, Y.-J.}, \bibinfo{author}{Oh,
  S.} \& \bibinfo{author}{Kim, J.}
\newblock \bibinfo{journal}{\bibinfo{title}{Quantum teleportation via a w
  state}}.
\newblock {\emph{\JournalTitle{New J. Phys.}}} \textbf{\bibinfo{volume}{5}},
  \bibinfo{pages}{136} (\bibinfo{year}{2003}).

\bibitem{LJK05}
\bibinfo{author}{Lee, S.}, \bibinfo{author}{Joo, J.} \& \bibinfo{author}{Kim,
  J.}
\newblock \bibinfo{journal}{\bibinfo{title}{Entanglement of three-qubit pure
  states in terms of teleportation capability}}.
\newblock {\emph{\JournalTitle{Phys. Rev. A}}} \textbf{\bibinfo{volume}{72}},
  \bibinfo{pages}{024302} (\bibinfo{year}{2005}).

\bibitem{HHH99}
\bibinfo{author}{Horodecki, M.}, \bibinfo{author}{Horodecki, P.} \&
  \bibinfo{author}{Horodecki, R.}
\newblock \bibinfo{journal}{\bibinfo{title}{General teleportation channel,
  singlet fraction, and quasidistillation}}.
\newblock {\emph{\JournalTitle{Phys. Rev. A}}} \textbf{\bibinfo{volume}{60}},
  \bibinfo{pages}{1888} (\bibinfo{year}{1999}).

\bibitem{BHH00}
\bibinfo{author}{Badziag, P.}, \bibinfo{author}{Horodecki, M.},
  \bibinfo{author}{Horodecki, P.} \& \bibinfo{author}{Horodecki, R.}
\newblock \bibinfo{journal}{\bibinfo{title}{Local environment can enhance
  fidelity of quantum teleportation}}.
\newblock {\emph{\JournalTitle{Phys. Rev. A}}} \textbf{\bibinfo{volume}{62}},
  \bibinfo{pages}{012311} (\bibinfo{year}{2000}).

\bibitem{MP95}
\bibinfo{author}{Massar, S.} \& \bibinfo{author}{Popescu, S.}
\newblock \bibinfo{journal}{\bibinfo{title}{Optimal extraction of information
  from finite quantum ensembles}}.
\newblock {\emph{\JournalTitle{Phys. Rev. Lett.}}}
  \textbf{\bibinfo{volume}{74}}, \bibinfo{pages}{1259} (\bibinfo{year}{1995}).

\bibitem{CKW00}
\bibinfo{author}{Coffman, V.}, \bibinfo{author}{Kundu, J.} \&
  \bibinfo{author}{Wootters, W.~K.}
\newblock \bibinfo{journal}{\bibinfo{title}{Distributed entanglement}}.
\newblock {\emph{\JournalTitle{Phys. Rev. A}}} \textbf{\bibinfo{volume}{61}},
  \bibinfo{pages}{052306} (\bibinfo{year}{2000}).

\bibitem{ZS22}
\bibinfo{author}{Zhou, L.} \& \bibinfo{author}{Sheng, Y.-B.}
\newblock \bibinfo{journal}{\bibinfo{title}{One-step device-independent quantum
  secure direct communication}}.
\newblock {\emph{\JournalTitle{Sci. China Phys. Mech. Astron.}}}
  \textbf{\bibinfo{volume}{65}} (\bibinfo{year}{2022}).

\bibitem{SBZY23}
\bibinfo{author}{Shi, W.-M.}, \bibinfo{author}{Bai, M.-X.},
  \bibinfo{author}{Zhou, Y.-H.} \& \bibinfo{author}{Yang, Y.-G.}
\newblock \bibinfo{journal}{\bibinfo{title}{Controlled quantum teleportation
  based on quantum walks}}.
\newblock {\emph{\JournalTitle{Quantum Information Processing}}}
  \textbf{\bibinfo{volume}{22}} (\bibinfo{year}{2023}).

\bibitem{AAC00}
\bibinfo{author}{Acín, A.} \emph{et~al.}
\newblock \bibinfo{journal}{\bibinfo{title}{Generalized schmidt decomposition
  and classification of three-quantum-bit states}}.
\newblock {\emph{\JournalTitle{Phys. Rev. Lett.}}}
  \textbf{\bibinfo{volume}{85}}, \bibinfo{pages}{1560} (\bibinfo{year}{2000}).

\bibitem{AJD00}
\bibinfo{author}{Acín, A.}, \bibinfo{author}{Jané, E.},
  \bibinfo{author}{Dür, W.} \& \bibinfo{author}{Vidal, G.}
\newblock \bibinfo{journal}{\bibinfo{title}{Optimal distillation of a
  greenberger-horne-zeilinger state}}.
\newblock {\emph{\JournalTitle{Phys. Rev. Lett.}}}
  \textbf{\bibinfo{volume}{85}}, \bibinfo{pages}{4811} (\bibinfo{year}{2000}).

\end{thebibliography}

\section*{Acknowledgements}
M.C. acknowledges support from the National Research Foundation (NRF) of Korea grant funded by the Korea Government
(Grant No. NRF-2022M3K2A1083890),
E.B. acknowledges support from the NRF of Korea grant funded by the Ministry of Science and ICT (MSIT) (Grant No. NRF-2022R1C1C2006396), 
and 
S.L. acknowledges support from 
the NRF grant funded by the MSIT (Grant No. NRF-2022R1F1A1068197), the MSIT, Korea, 
under the Information Technology Research Center support program (Grant No. IITP-2023-2018-0-01402) 
supervised by the Institute for Information and Communications Technology Planning and Evaluation and the Creation of the Quantum Information Science R$\&$D Ecosystem (Grant No. 2022M3H3A106307411) through the NRF funded by the MSIT.

\section*{Author contributions statement}

M.C., E.B. and S.L. conceived the idea. M.C. performed the calculations and the proofs, and E.B. and S.L. checked them. M.C. wrote the main manuscript, and E.B. and S.L. improved the manuscript. All authors contributed to the discussion and
reviewed the manuscript.

\section*{Competing interests}
The authors declare no competing interests.

\end{document}